\documentclass[
aps,%
12pt,%
final,%
notitlepage,%
oneside,%
onecolumn,%
nobibnotes,%
nofootinbib,% In the current version of REVTeX, when this option is on,
superscriptaddress,%
noshowpacs,%
centertags]%
{revtex4}
\begin{document}
%\selectlanguage{russian} % For a paper in Russian
\selectlanguage{english} % For a paper in English

\hyphenation{fourier-trans-formed sin-gle di-a-gram anoma-lous mag-netic mo-ment}

\title{Effects of perturbative exchanges in a QCD--string model}

\date{\today}

\author{J. Weda}
\affiliation{Institute for Theoretical Physics, University of Utrecht, The Netherlands} 
\author{J. A. Tjon}
\affiliation{Jefferson Lab, 12000 Jefferson Ave,  Newport News, VA 23606, USA}

\begin{abstract}
The QCD-string model for baryons derived by Simonov
and used for the calculation of baryon magnetic moments in 
Ref.~\cite{Simonov:2001iz} is extended
to include also perturbative gluon and meson exchanges. The mass
spectrum of the baryon multiplet is studied. For the meson 
interaction either the pseudoscalar or pseudovector coupling is used.
Predictions are compared with the experimental data.
Besides these exchanges the influence of excited quark orbitals on the 
baryon ground state are considered by performing a multichannel calculation.
The nucleon-$\Delta$ 
splitting increases due to the mixing of higher quark states while the
baryon magnetic momenta decrease. The multichannel calculation with
perturbative exchanges is shown to yield reasonable magnetic moments while
the mass spectrum is close to experiment.
\end{abstract}

\maketitle

\section{Introduction}

As QCD is generally accepted as the theory of strong interactions,
$\bar{q}q$ and $3q$ dynamics should be derived from QCD in a fully 
relativistically covariant way. This is a 
formidable task due to the large gluon-quark
coupling constant in the low energy regime and the non-Abelian 
character of QCD. 
In recent years the formalism of field correlators was set up to deal 
with the two main characteristics of quark-systems: confinement and
chiral symmetry breaking. Using this field correlator method (FCM)
a nonlinear equation has been derived for a light quark in the field
of a heavy antiquark \cite{sim9,sim10}. In the derivation use has been
made of the large $N_c$ limit and the calculation has been restricted
to include only the Gaussian (bilocal) field correlator, an approximation
which has been shown to be correct within a few percent by lattice 
simulations as was discussed in Ref.~\cite{shesim}.

The developed method is quite general and can be extended to treat 
the light quark systems $\bar{q}q$ and $3q$ as was shown in 
Refs.~\cite{Simonov:2001iz,sim12}.
Using this formalism the baryon magnetic moments and corrections on 
these from virtual mesonic excitations have been calculated. Reasonable
agreement with experiment has been obtained without the need of introduction 
of constituent quark masses~\cite{Simonov:2001iz}.
The outcome of the method is partially summarized in the next section.  

In the present paper we extend the model in the following way. 
Using the same formalism as in Ref.~\cite{Simonov:2001iz} to obtain 
the baryon wave functions, the baryon mass spectrum of the lowest octet and
decuplet representation of the $SU(3)$-flavor group is obtained. 
In section \ref{sec:exchangepotentials} perturbative corrections to 
this spectrum due to one gluon (OGE) and pseudoscalar meson exchanges
are calculated. 
For the coupling of the pseudoscalar mesons to the quark both pseudoscalar and 
pseudovector couplings are exploited. 
Following Refs.~\cite{Isgur:1978xj,Isgur:1979wd,Capstick:1986bm} we 
consider the influence of excited states to the ground state. This is
done by forming excited baryon states out of the orbital and radial 
excitations of the single-quark wave functions and using these as a 
basis to diagonalize the Hamiltonian. The results from the multichannel 
calculation are shown in section~\ref{sec:mccalculation}.
The consequences of the diagonalization of the Hamiltonian to the 
magnetic moments are calculated in section~\ref{sec:magnmom}.
The paper ends with some discussions in section~\ref{sec:conclusion}.

\section{Formalism}

As was explained in Refs.~\cite{Simonov:2001iz,sim12} the field correlator
method can be used to obtain an effective quark Lagrangian $L_{eff}$ from
the QCD-partition function. Integrating out the gluon fields 
by using the generalized Fock-Schwinger gauge with contour 
$C$~\cite{shesim2,Ivanov:1985bk,Ivanov:1986np} 
the QCD action can be rewritten as
\be
Z= \int D\kappa(C) D\Psi D\Psi^{\dagger} exp~({L_0+L_{eff}}),
\label{eq1}
\ee
where $L_0$ is the free quark field Lagrangian and the effective Lagrangian in 
lowest order in quark fields has the form of a nonlocal four-fermion interaction.
Moreover, since $Z$ is gauge invariant we have introduced an additional integration over a set of contours $C(k)$ with weight $D\kappa(C)$ in the partition function. 
In the field correlator method the Gaussian approximation is usually made, 
which is has been discussed extensively in
Refs.~\cite{shesim,sim11}. Choosing the generalized Fock-Schwinger 
gauge~\cite{shesim2,Ivanov:1985bk,Ivanov:1986np} with
contours going through a given point ${\bf r_0}$ an effective action is determined. 
In this way a Hamiltonian 
equation has  been derived for the baryon, depending on the parameter ${\bf r_0}$,
\be
(H_1+H_2+H_3-E)\Psi(
\mbox{\boldmath
  ${\rm r}$}_1,
\mbox{\boldmath
  ${\rm r}$}_2,
\mbox{\boldmath
  ${\rm r}$}_3)=0,
\label{eq2}
\ee
with
\be
H_i=m_i\beta^{(i)}+
\mbox{\boldmath ${\rm  p}$}_i
\mbox{\boldmath ${\rm  \alpha}$}^{(i)}+
\beta^{(i)}M_i(
\mbox{\boldmath ${\rm  r}$}_i-
\mbox{\boldmath ${\rm r} $}_0).
\label{eq3}
\ee
The quark mass operator $M_i$ produces both linear confinement and
chiral symmetry breaking as was shown in Refs.~\cite{simtjo2,sim9,sim10}.
It is a nonlinear and nonlocal operator acting in the
coordinate space of the $i$-th quark.
When only the dominant part of $M_i$ leading
to confinement is kept, the kernel can be characterized as
\be
M(x,y)\approx~ \frac{1}{2T_g\sqrt{\pi}}
\sigma\left|\frac{\mbox{\boldmath ${\rm x}$}+\mbox{\boldmath ${\rm y}$}}{2}\right|
\exp\left(-\frac{({\bf x}-{\bf y})^2}{4T_g^2}\right),
\label{eq4}
\ee
where $T_g$ is the gluon correlation length corresponding
to the length scale of correlations in the fluctuations of the 
gluon background field. From lattice gauge simulations 
it has been found to be of the order of $\frac{1}{4}~fm$. 
Following Ref.~\cite{simtjo2} we have adopted a value 
$T_g=0.24~fm$. For asymptotic large $|x|$ it has been shown 
that the kernel Eq.~(\ref{eq4}) leads for a fixed ${\bf r_0}$ 
to a linear confining interaction $\sigma |{\bf r_i}-{\bf r_0}|$ 
\cite{simtjo2,simtjo3}. 
Moreover, it is of a Lorentz-scalar type. In this paper we will
also allow for a constant term in the confining interaction, 
corresponding to the next leading order corrections to the 
area law. 
The weight $D\kappa(c)$ should be chosen as a stationary point of the effective action 
and that the contours generate a string of minimal length. 
As a consequence the parameter $\ver_0$ can in principle 
be found as the minimum of the interaction between the three quarks, 
yielding for $\ver_0$ the so-called Torricelli point. 
This would result in a single string Y-junction which is of a
genuine three-body nature.
However, for practical reasons, we take as a first approximation
$\ver_0$ as a constant parameter as was done in 
Ref.~\cite{Simonov:2001iz}.
Due to treating $\ver_0$ as a constant instead of adopting the Torricelli 
point $\ver_T$, the distance $\sum_{i=1}^3\left|\ver_i-\ver_0\right|$ is
about $\sim 2.5$ times bigger than the minimal distance  
$\sum_{i=1}^3\left|\ver_i-\ver_T\right|$. This means that the string
tension $\sigma$ has to be chosen $\sim 2.5$ times smaller to yield
about the same energy. Recent lattice simulations on static quarks obtain 
a value of about $\sigma=0.15~\text{GeV}^2$ for the three-quark Y-shaped
interaction which is close to the value for quark-antiquark 
interactions~\cite{Takahashi:2000te}.
This result suggests that string tensions as low as  $\sigma=0.06~\text{GeV}^2$
can be used in our calculations.

The solution of Eq.~(\ref{eq2}) is now of a factorizable form which
enables us to represent the baryon wave function as the
product of three single particle solutions,
\be
\Psi_{JM} = \Gamma^{\alpha\beta\gamma}_{JM} (f_1f_2f_3) e_{abc}
 \psi^{f_1}_{a\alpha}(\ver_1-\ver_0) 
 \psi^{f_2}_{b\beta}(\ver_2-\ver_0) 
 \psi^{f_3}_{c\gamma}(\ver_3-\ver_0),
\label{eq5}
\ee
where $a,~b,~c$ and $f_i$ are the color and flavor indices respectively.
The orbital and radial excitations are indicated by $\alpha$, $\beta$
and $\gamma$. 
Because of the Pauli principle the baryon wave function has to be 
a total antisymmetric function of the three quark functions. As the
baryon is a color white object (antisymmetric in color), the 
wave function Eq.~(\ref{eq5}) has to be total symmetric in flavor,
orbital and radial excitations of the single particle wave functions
$\psi^{f_i}_{a\alpha}$. The functions $e_{abc}$ and  
$\Gamma^{\alpha\beta\gamma}_{JM}$ respectively take care of this 
(anti)-symmetrization.
Explicit formulas for the lowest baryons are given in Appendix~\ref{app:sym}.

Each single particle solution satisfies a nonlinear Dirac-like 
Hamiltonian equation,
\be
H_i\psi^{f_i}_{a_i\alpha_i}=\varepsilon_{n_i}^{(i)}\psi^{f_i}_{a_i\alpha_i},
\qquad E=\sum_{i=1}^3 \varepsilon_{n_i}^{(i)}.
\label{eq6}
\ee
where $H_i$ is given by Eq.~(\ref{eq3}). It has to be
solved self consistently, leading to confining
solutions. Details can be found in Ref.~\cite{simtjo2}. 
In table~\ref{tab:solution} some values of $\varepsilon_n$ are shown.
The solutions are listed as
\be
n(j)^{{\rm sign}(\kappa){\rm sign}(\varepsilon_n)},
\qquad
\kappa=\pm\left(j+\tfrac12\right).
\label{eq7}
\ee
From these single-quark orbitals the baryon spectrum 
$M_B=E$ and the corresponding baryon wave funtions, 
Eq.~(\ref{eq5}), 
can immediately be constructed.

Using these baryon wave functions the magnetic moments $\mu_z$ 
have been determined in Ref.~\cite{Simonov:2001iz}.
Following Ref.~\cite{klo} we have
\be
\mu_{z} = \frac{e}{2M_p} G_{mag}(Q=0) = 
-\frac{i}{2} \left[ \venab_{Q} \times \veM \right]_{z}({Q=0}),
\label{eq8}
\ee
where $e$, $M_p$ and $G_{mag}$ are the charge, mass and
Sachs magnetic moment of the proton.
The electromagnetic current matrix element $M_{\mu}$ is
given by
\be
M_{\mu}= \left<\Psi\left| J_{\mu}(Q)\right|\Psi\right>.
\label{eq9}
\ee
For the operator $J_\mu$ the single-quark current operator is 
taken as the first approximation. Higher order contributions 
come from two-body currents like one-pion-exchange currents 
and mesonic one-loop corrections which give rise to the 
anomalous magnetic moment of the quark. 
Reasonable values were obtained for the magnetic moments of 
the baryon octet and decuplet in the case of a string tension 
of $\sigma=0.1~\text{GeV}^2$ without the need of introduction 
of constituent quark masses~\cite{Simonov:2001iz}. 
In the present paper we extend this analysis to
also include the one-gluon-exchange and study the
effects of the use of different forms
for the interaction in the one-pion-exchange.

\section{Exchange potentials}
\label{sec:exchangepotentials}

\begin{figure*}[t!]
\setcaptionmargin{5mm}
\onelinecaptionsfalse
\includegraphics[width=8cm]{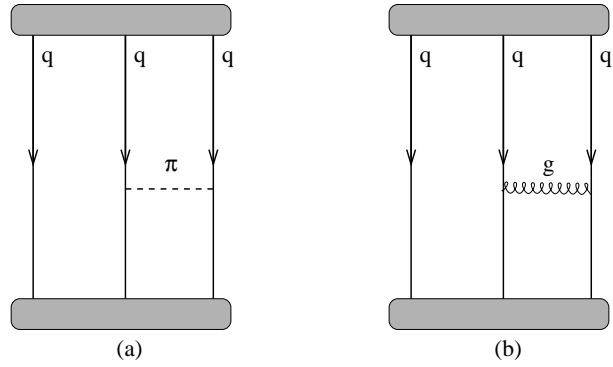}
\captionstyle{normal}
\caption{ A schematic view of the exchange of a
pion (a) and a gluon (b) between two quarks in a baryon. The blobs at the
beginning and the end of the diagram represent bound states of the quarks.
\label{fig:effpion}}
\end{figure*}

Until now the picture of quarks moving in the confining sea of gluons
is used. This leads to confinement and chiral symmetry breaking, 
but it lacks spin-dependent interactions. Typical spin-dependent  
structures in the baryon spectrum as the splitting between the 
nucleon and the $\Delta$ are therefore not present in such a
simplified model.  To remove this deficiency a perturbative 
one-gluon-exchange interaction is now introduced. The Coulomb 
part is expected to lower the baryon masses and the color 
magnetic part is expected to give rise to a splitting between 
the $J=1/2$ and $J=3/2$ states.
Beside this one-gluon interaction we also introduce perturbative 
pseudoscalar meson-exchanges which can be considered as 
an effective interaction representing 
the exchange of correlated quark anti-quark pairs.
The interactions are schematically shown in 
Figure~\ref{fig:effpion}.

The Hamiltonian equation, Eq.~(\ref{eq2}), is equivalent
to a Bethe-Salpeter equation with an instantaneous
interaction. Of all possible three particle wave functions of the 
type (\ref{eq5}) it couples only to $(+++)$ and
$(---)$, where $\pm$ indicates the sign of 
$\epsilon_n$ as in Eq.~(\ref{eq7}). All other wave functions decouple.
Within such an equal time approximation we now consider
the effects of the perturbative exchanges.
Following Refs.~\cite{Loring:2001kv,Metsch:1997bf,Metsch:2001bs}
where the instantaneous three particle Bethe-Salpeter 
equation is considered, we assume that the perturbative
gluon- and meson-exchanges
only take place between these purely 
positive and negative energy components. 
As discussed in Ref.~\cite{Loring:2001kv} we write for the 
interaction potential in the Hamiltonian,
\begin{multline}\label{eq10}
H_{\rm int}^{(23)}\Psi(P,\vep'_{\rho},\vep'_{\lambda})
=\left(\Lambda^+(\vep_1)\Lambda^+(\vep_2)\Lambda^+(\vep_3)+\Lambda^-(\vep_1)\Lambda^-(\vep_2)\Lambda^-(\vep_3)
\right)\\
\times\int\frac{d^3 p'_{\rho}}{(2\pi)^3}\frac{d^3 p'_{\lambda}}{(2\pi)^3}\gamma^0_{(2)}\gamma^0_{(3)}V_{\rm ex}^{(23)}(\vep_{\rho},\vep_{\lambda},\vep'_{\rho},\vep'_{\lambda})\Psi(P,\vep'_{\rho},\vep'_{\lambda}),
\end{multline}
with the energy projection operators defined as
\be
\Lambda^{\pm}(\vep)=\frac{\omega\pm H(\vep)}{2\omega},
\qquad 
\omega=\left|\epsilon_n\right|.
\label{eq11}
\ee
The interactions between the other quark pairs, $H_{\rm int}^{(12)}$
and $H_{\rm int}^{(13)}$, can very similarly be written down and
easily be included. The Hamiltonian equation Eq.~(\ref{eq2}) becomes
\begin{align}
H\Psi(\ver_1,\ver_2,\ver_3)&=
\left(H_1+H_2+H_3+
H_{\rm int}^{(12)}+H_{\rm int}^{(13)}+H_{\rm int}^{(23)}\right)
\Psi(\ver_1,\ver_2,\ver_3)
\notag\\
&=\left(E+\Delta E\right)\Psi(\ver_1,\ver_2,\ver_3).
\label{eq12}
\end{align}
The exchange interactions are treated perturbatively,
\be
\Delta E=3\left\langle\Psi_{JM}
\left|H_{\rm int}^{(23)}\right|
\Psi_{JM}\right\rangle,
\label{eq13}
\ee
where $\Psi_{JM}$ represents the baryon wave function, Eq.~(\ref{eq5}).
Because of symmetry considerations the interactions
between the other quark pairs can simply be included by adding a factor of 
3 shown in Eq.~(\ref{eq13}).

The exchange potential $V_{\rm ex}^{(23)}$ in Eq.~(\ref{eq10}) for the
non-strange baryons can 
be split into two contributions, the one gluon exchange (OGE) 
and the one-pion-exchange (OPE). They are explicitly written as,
\begin{subequations}
\label{eq14}
\begin{align}
V_{oge}^{(23)}(\vek) & =  -4\pi\frac23\alpha_s\frac{\gamma_{\mu}(2)\gamma^{\mu}(3)}{\vek^2}, 
\label{eq14a}\\
V_{ope}^{(23)}(\vek) & =  4\pi  g_{qq\pi}^2 \vetau(2) . \vetau(3)\gamma_5(2)\gamma_5(3)\frac{1}{\vek^2+m_{\pi}^2}\left(\frac{\Lambda^2}{\vek^2+\Lambda^2}\right)^2, 
%\qquad\text{(PS)},
\label{eq14b}
\end{align}
\end{subequations}
where a pseudoscalar (PS) coupling for the pion is assumed. 
The factor $2/3$ in the OGE originates from the color 
content of the baryons and the term $\vetau(2) . \vetau(3)$
in the OPE takes care of the isospin.
The exchange potentials can also be written in coordinate space
by performing Fourier transformations
\begin{subequations}
\label{eq15}
\begin{align}
V_{oge}^{(23)}(\ver_2-\ver_3) & =  -\frac23\alpha_s\frac{\gamma_{\mu}(2)\gamma^{\mu}(3)}{\left|\ver_2-\ver_3\right|},
\label{eq15a}\\
V_{ope}^{(23)}(\ver_2-\ver_3) & = 
g_{qq\pi}^2 \vetau(2) . \vetau(3)\gamma_5(2)\gamma_5(3) \left(\frac{\Lambda^2}{\Lambda^2-m_{\pi}^2}\right)^2
\label{eq15b}\\ 
&\quad\times
\left(-\frac{e^{-m_{\pi}\left|\ver_2-\ver_3\right|}}{\left|\ver_2-\ver_3\right|}+
\frac{e^{-\Lambda\left|\ver_2-\ver_3\right|}}{\left|\ver_2-\ver_3\right|}+
\frac{\Lambda^2-m_{\pi}^2}{\Lambda}
\frac{e^{-\Lambda\left|\ver_2-\ver_3\right|}}{2}\right),
\qquad\text{(PS)}.
\nonumber
\end{align}
\end{subequations}
Details on the calculation
of the perturbative calculation of the matrices Eq.~(\ref{eq13})
can be found in appendix~\ref{app:exchpot}.

When a pseudovector (PV) pion-quark coupling is assumed some 
modifications have to be made. In Eq.~(\ref{eq14b}) the
pseudovector coupling is obtained by the replacement
\be
g_{\pi qq}\gamma_5\tau_a
\rightarrow
\frac{g_{\pi qq}}{2 m_{eff}}\gamma_5\not\! k\tau^a,
\label{eq16}
\ee
where $m_{eff}$ is a scaling mass of the order of the constituent 
quark mass, to be discussed later.
By Fourier transformation the momenta in Eq.~(\ref{eq16}) 
turn into derivatives in coordinate space.
In this way the matrices $\not\! k$ can be evaluated when they act 
upon the quark wave functions resulting in some derivatives
on the radial wave function and a changed angular dependence.
Details are given in Appendix~\ref{app:exchpot}.

The model can easily be extended to include the strange baryons. 
Apart from pion exchanges we now also have to add kaon and eta exchanges. 
These exchanges can be included by changing the SU(2) isospin matrices
$\vetau_2 . \vetau_3$ in Eq.~(\ref{eq14b}) into SU(3) flavor matrices
$\sum_{a=1}^{8}\lambda_2^a\lambda_3^a$ and substituting the correct
coupling constants, meson and cutoff masses.

For the coupling of the gluon to the quark a running coupling constant 
is used which depends on the distance. 
Following Ref.~\cite{Badalian:2001by} the coupling is parameterized as
\be
\alpha^{(2)}(q)=\alpha^{(1)}(q)
\left(1-\frac{\beta_1}{\beta_0^2}\frac{\ln t_B}{t_B}\right),
\qquad
\alpha^{(1)}(q)=\frac{4\pi}{\beta_0 t_B},
\label{eq17}
\ee
with
\be
t_B=\ln\frac{q^2+m_B^2}{\Lambda_V^2},
\ee
and
\be
\beta_0=11-\frac23 n_f,\qquad 
\beta_1=102-\frac{38}{3}n_f.
\ee
As the calculation is performed in coordinate space, the coupling constant 
has to be Fouriertransformed which can be done as,
\be
\tilde{\alpha}(r)=\frac2\pi\int_0^{\infty} dq \frac{\sin(qr)}{q}\alpha(q)
=\frac2\pi\int_0^{\infty} dx \frac{\sin x}{x}\alpha(x/r).
\ee
The constants are fixed at $m_B=1.0~GeV$, $\Lambda_V=385~MeV$ and
$n_f=3$ as discussed in Ref.~\cite{Badalian:2001by}.

Following Ref.~\cite{glo} we exploit the Goldberger-Treiman relation 
\cite{Goldberger:1958tr} for the 
pion-quark vertex to find the pseudovector coupling constant to meson $m$
\be
f_{mqq}=M_m\frac{g_A^q}{2f_m},
\label{eq18}
\ee
where $M_m$ is the meson mass, $g_A^q$ is the quark axial coupling constant
and $f_m$ is the decay constant. From Refs.~\cite{Lahde:2002fe,PDBook} we take
the decay constants $f_\pi=93$~MeV, $f_\kappa=113$~MeV and $f_\eta=112$~MeV.

The axial coupling constant of the quark is not well known. In the static 
quark model it can be related to the nucleon axial coupling constant as, 
$g_A^q=\tfrac35 g_A^N$, which would give $g_A^q=0.75$ for $g_A^N=1.26$. 
In the large $N_c$ limit however the coupling would be $g_A^q=1$ 
as was derived in Ref.~\cite{Weinberg:1990xm} and confirmed by Simonov
using the field correlator method in Ref.~\cite{Simonov:2002er}.
If $1/N_c$ corrections are taken into account the coupling decreases
to $g_A^q=0.87$ \cite{Weinberg:1991gf}. 
For consistency reasons we try to keep close to the field correlator 
method and choose $g_A^q=1$.
The resulting values for $f_{mqq}$ are determined from the Goldberger-Treiman
relation and shown in Table~\ref{tab:parameters}.
The parameterization of the cutoff mass is taken the same as in Ref.~\cite{glo}
\be
\Lambda_m=\Lambda_0+\kappa M_m
\ee
with the parameters $\Lambda_0=565$~MeV, $\kappa=0.81$, the results
are given in Table~\ref{tab:parameters}.

The pseudoscalar coupling constant $g_{mqq}$ is related to the pseudovector 
coupling constant $f_{mqq}$ as,
\be
\frac{g_{mqq}}{2m_{eff}}=\frac{f_{mqq}}{M_m}.
\ee
The $m_{eff}$ can be looked at as the effective constituent mass of the quark.
In case of the pion-coupling $m_{eff}$ is the effective constituent mass of the 
$u,d$-quark, in case of the kaon- and eta-coupling $m_{eff}$ is a mixture
of $u$-, $d$- and $s$-quark masses.
The mass is chosen such that the one-pion and one-kaon exchanges using PS coupling 
give the same value as using the PV coupling in only positive energy channels. 
So we require
\be
V^{(23)}_{ome}(PS)=V^{(23)}_{ome}(PV)\left(\frac{1+\gamma_0}{2}\right)_{(2)}
\left(\frac{1+\gamma_0}{2}\right)_{(3)},
\qquad m=\pi,\kappa.
\label{eq19}
\ee
The effective mass $m_{eff}$ for the eta meson is put equal to the $m_{eff}$ for
the kaon meson.
The parameters are summarized in Table~\ref{tab:parameters}. It can be seen that
the coupling constants $g_{mqq}$ are almost equal, as would be the case in
a chiral symmetric world. In case of $\sigma=0.09~\text{GeV}^2$ we find values 
which are somewhat smaller than $g_{mqq}^2/4\pi=0.67$ as was used by 
Glozman et al.~\cite{Glozman:1996fu,glo}, in
case of $\sigma=0.12~\text{GeV}^2$ somewhat larger.

Using these coupling constants the perturbative exchanges are calculated, 
where as a first approximation the baryon wave function Eq.~(\ref{eq5})
is used.
Results are shown in Tables~\ref{tab:spectrum0}, \ref{tab:spectrum1} and
\ref{tab:spectrum1b} for string tensions of $\sigma=0.06~\text{GeV}^2$,
$\sigma=0.09~\text{GeV}^2$ and $\sigma=0.12~\text{GeV}^2$ respectively.

As can be seen from Tables~\ref{tab:spectrum0}, \ref{tab:spectrum1} 
and \ref{tab:spectrum1b} an extra parameter $C_0$ has 
been introduced which is of a Lorentz-vector nature. 
This parameter is added to the confining potential as
\be
M(\ver^{(i)}-\ver^{(0)})\rightarrow
M(\ver^{(i)}-\ver^{(0)})+\beta^{(i)}C_0.
\label{eq20}
\ee
In the derivation of the confining potential in Refs.~\cite{sim12,simtjo2} 
the large distance behavior of the interaction was examined. The actual 
dependence at short distances however is quite unknown. There can be 
contributions of the Lorentz-vector or scalar type, even spin--spin 
interactions are possible. Our results seem to suggest that a constant 
$C_0$ with a value of about $170-190$~MeV has to be added in case of the 
PS-coupling.

When the PV-coupling is used we find smaller values of the meson exchanges 
as compared to the PS-coupling. Considering only positive energy components 
both couplings give the same results by definition, Eq.~(\ref{eq19}). 
The inclusion of negative energy components decreases the effect
of meson exchanges such that a smaller value of $C_0$, in the range of  
$C_0\approx140-160$~MeV, is needed in case of a PV-coupling.

The hyperfine splitting and the OPE show some symmetry. The origin of the
symmetry can be understood by performing a 
nonrelativistic reduction on the exchanges Eqs.~(\ref{eq14}). 
The resulting Breit interaction yields the $\vesig_2 . \vesig_3$ structure 
which causes the well known $-3:1$ splitting between the spin singlet and 
triplet state.
From Appendix~\ref{app:sym} it can be seen that the $\Delta$ solely consists
of spin triplet ($j_{23}=1$) states and the nucleon is build up from a sum 
of spin singlet ($j_{23}=0$) and spin triplet ($j_{23}=1$) states. This 
results exactly in the $-1:1$ hyperfine splitting observed for the nucleon 
and the $\Delta$ in Tables~\ref{tab:spectrum0}, \ref{tab:spectrum1} and 
\ref{tab:spectrum1b}. 
The splitting $-5:-1$ in the OPE can similarly be understood when it is 
realized that in the nonrelativistic reduction of Eq.~(\ref{eq15b}) the 
spin-isospin structure looks like $\vesig_2 . \vesig_3~~\vetau_2
. \vetau_3$.

When the string tension $\sigma$ is increased the single particle orbitals
tend to become more compact. Due to the $1/r$ behavior of the exchange
potentials this results in larger values in the perturbative calculation
Eq.~(\ref{eq13}).

The results for the 
baryon octet ($J=1/2$) are quite close to the experimental masses in 
case of the PS-coupling for $\sigma=0.06~\text{GeV}^2$ and in case of 
the PV-coupling for $\sigma=0.09~\text{GeV}^2$.
From Table~\ref{tab:spectrum1} it can be seen that 
the PS-coupling does somewhat better for the 
baryon decuplet ($J=3/2$) where the PV-calculation misses about 100~MeV.
For $\sigma=0.12~\text{GeV}^2$ in Table~\ref{tab:spectrum1b} the PV-coupling
leads to a reasonable overall agreement while the PS-calculation produces
values which are too large.

\section{Multichannel calculation}
\label{sec:mccalculation}

Until this far the baryon wave functions 
contain only the ground state of the single-quark
orbital. That is, the baryon wave function Eq.~(\ref{eq5}) can 
schematically be written as
\be
\Psi_{JM}=\prod_{k=1}^{3}\left(0\left(\tfrac12\right)^{++}\right)_k,
\label{eq21}
\ee
where the notation from Eq.~(\ref{eq7}) has been used.
Coefficients needed for symmetrization and coupling to the proper
angular momentum are left out for simplicity. 
Quarks however can also be in excited orbitals which means that generally
baryon wave functions which are (partly) build up from excited single 
particle solutions also contribute to the baryon. So contributions like
\be
\begin{aligned}
\Psi_{JM}&=\left(1\left(\tfrac12\right)^{++}\right)_1
\prod_{k=2}^{3}\left(0\left(\tfrac12\right)^{++}\right)_k,
\\
\Psi_{JM}&=\left(0\left(\tfrac12\right)^{++}\right)_1
\left(0\left(\tfrac12\right)^{-+}\right)_2
\left(1\left(\tfrac12\right)^{++}\right)_3,
\\
\Psi_{JM}&=\left(0\left(\tfrac32\right)^{++}\right)_1
\left(0\left(\tfrac12\right)^{++}\right)_2
\left(0\left(\tfrac12\right)^{-+}\right)_3,
\end{aligned}
\label{eq22}
\ee
can mix into the baryon ground state and change the energy.
Similar as was done in Refs.~\cite{Isgur:1978xj,Isgur:1979wd} for 
nonrelativistic quark models and in Ref.~\cite{Capstick:1986bm} for
a relativized model we take
wave functions as Eqs.~(\ref{eq22}) as a basis for diagonalizing the
Hamiltonian Eq.~(\ref{eq12}). 

As the color content of the baryon takes care of the antisymmetrization
the resulting \mbox{3-particle} wave function has to be total symmetric with 
respect to particle interchanges when the color is disregarded.
Spin, isospin, orbital and radial excitations are taken into account in
the symmetrization procedure. The 3-quark state can be characterized in
the following way. Let us consider the representation, where
quark 1 plays a special role. Starting from the single-quark orbitals
we may couple quark 2 and 3 to a  $j_{23}$ and $i_{23}$ state. 
These states can in general be made symmetric by adding the permutation 
$2\leftrightarrow 3$ to these states. 
An appropriate choice for the quantum numbers $j_{23}$ and $i_{23}$ 
has to be made such that the wave function does not vanish.
To form the 3-quark state with total quantum numbers J and I, the single-quark
orbital of quark 1 is added.
Taking the proper linear combinations of the Faddeev components formed from the first term
by the interchanges $1\leftrightarrow2$ and $1\leftrightarrow3$ it is assured that the whole
wave function is total symmetric under the interchange of any two particles.
Adopting for the moment the notation
\be
\left|n_kj_kl_km_{j_k}\right>\otimes\left|i_km_{i_k}\right>,
\ee
to represent the single orbital solution we have
\begin{multline}
\Psi_{J,I}(j_{23},i_{23})=
\sum \text{C.G.}\:
\Bigl\{
\left|1\right>
\bigl(\left|2\right>\left|3\right>+\left|3\right>\left|2\right>\bigr)
+\left(1\leftrightarrow2\right)+\left(1\leftrightarrow3\right)
\Bigr\}
\displaybreak[0] \\
\shoveleft{=
\sum_{\text{$m_{i}$'s, $m_{j}$'s}}
C(i_1,i_{23},I;m_{i_1},m_{i_{23}},M_I)
C(j_1,j_{23},J;m_{j_1},m_{j_{23}},M_J)}
\displaybreak[0] \\
\shoveright{\times
C(i_2,i_3,i_{23};m_{i_2},m_{i_3},m_{i_{23}})
C(j_2,j_3,j_{23},m_{j_2},m_{j_3},m_{j_{23}})}
\displaybreak[0] \\
\shoveleft{\times\biggl\{
\left|n_1j_1l_1m_{j_1}\right>
\otimes
\left|i_1m_{i_1}\right>
\Bigl(
\left|n_2j_2l_2m_{j_2}\right>\left|n_3j_3l_3m_{j_3}\right>
\otimes
\left|i_2m_{i_2}\right>\left|i_3m_{i_3}\right>
\Bigr.+\biggr.}
\displaybreak[0] \\
\biggl.\Bigl.
\left|n_3j_3l_3m_{j_2}\right>\left|n_2j_2l_2m_{j_3}\right>
\otimes
\left|i_3m_{i_3}\right>\left|i_2m_{i_2}\right>
\Bigr)
+\left(1\leftrightarrow2\right)+\left(1\leftrightarrow3\right)
\biggr\},
\label{eq23}
\end{multline}
where the  $C$'s are the Clebsch-Gordon coefficients in the Rose notation~\cite{rose}.
An allowed choice of  $j_{23}$ and $i_{23}$ does not always lead to unique 3-particle
wave functions. 
If one takes for example three identical single-particle orbitals with 
$j_1=j_2=j_3=1/2$ and $i_1=i_2=i_3=1/2$ the wave function 
$\Psi_{1/2,1/2}(0,0)$ equals $\Psi_{1/2,1/2}(1,1)$. 
Independence of these basis functions is tested 
by calculating the determinant of the matrix 
$\bigl<\Psi_\alpha\bigl|\Psi_\beta\bigr>$ where $\Psi_{\alpha}$,  
$\Psi_{\beta}$ are wave functions like Eq.~(\ref{eq23}).
A simple example of this procedure is shown in Appendix~\ref{app:sym}
where the total symmetric nucleon- and $\Delta$-wave functions are 
constructed.

The wave functions $\Psi_{\alpha}$ 
are taken as a basis to solve the full Hamiltonian, including 
one-gluon and one-pion exchange.
So let us consider
\be
\lan \Psi_{\alpha}|V_{\text{ex}}|\Psi_{\beta}\ran + \lan \Psi_{\alpha}| [H_0 - E] |\Psi_{\beta}\ran =0
\ee
with
\be
H_0|\Psi_{\beta}\ran= E_{\beta} |\Psi_{\beta}\ran,
\qquad
V_{\alpha\beta} = \lan\Psi_{\alpha}|V_{\text{ex}}|\Psi_{\beta}\ran,
\qquad
A_{\alpha\beta}= \lan\Psi_{\alpha}|\Psi_{\beta}\ran,
\ee
and $E_{\beta}=\epsilon^{(1)}+\epsilon^{(2)}+\epsilon^{(3)}$, 
the equation becomes,
\be
\left[A^{-1}_{\alpha\gamma} V_{\gamma\beta} +E_{\beta}\delta_{\alpha\beta}\right] = H_{\alpha\beta},
\ee
from which the eigenvalues have to be found.

From the calculations performed it is found that
the single-quark orbitals $1\bigl(\tfrac12\bigr)^{++}$, 
$0\bigl(\tfrac32\bigr)^{++}$ and to a smaller extend 
$0\bigl(\tfrac12\bigr)^{-+}$ give the largest change of the spectrum 
in case of a PS-coupling.
These orbitals have been used in a multichannel calculation for both the
$u$, $d$-quark and the $s$-quark. That is, all possible combinations of 
$0\bigl(\tfrac12\bigr)^{++}(u,d)$, $1\bigl(\tfrac12\bigr)^{++}(u,d)$, 
$0\bigl(\tfrac32\bigr)^{++}(u,d)$, $0\bigl(\tfrac12\bigr)^{-+}(u,d)$,
$0\bigl(\tfrac12\bigr)^{++}(s)$, $1\bigl(\tfrac12\bigr)^{++}(s)$, 
$0\bigl(\tfrac32\bigr)^{++}(s)$ and $0\bigl(\tfrac12\bigr)^{-+}(s)$ which
couple to some specific total angular momentum $J$ and total isospin $I$ are
taken into account.
When a PV-coupling is exploited the single-particle orbital 
$0\bigl(\tfrac52\bigr)^{++}$ is taken instead of $1\bigl(\tfrac12\bigr)^{++}$ 
as it was found to give a larger contribution in this case.
The results are shown in Tables~\ref{result4channel0}, \ref{result4channel} 
and \ref{result4channelb} for string tensions of $\sigma=0.06~\text{GeV}^2$, 
$\sigma=0.09~\text{GeV}^2$ and $\sigma=0.12~\text{GeV}^2$ respectively.

The inclusion of the excited quark orbitals changes the spectrum 
considerably. All ground state masses lower through this calculation.
An effect which is stronger for the baryon octet, which 
causes the nucleon-$\Delta$ splitting to increase by about 100~MeV.
In case of $\sigma=0.09~\text{GeV}^2$ the situation for 
the baryon decuplet thus improves considerably, leading to a rather 
close prediction for the PV-calculation as can be seen in 
Table~\ref{result4channel}. 
The baryon octet however is quite well reproduced in 
Table~\ref{result4channel0} for a string tension of 
$\sigma=0.06~\text{GeV}^2$ and a PV-coupling.
The PS-calculation with the string tension of 
$\sigma=0.09~\text{GeV}^2$ yields values somewhat too large,
while the results in Table~\ref{result4channelb} are much too 
high for the decuplet.

As a second consequence of the lower masses, larger $C_0$ values have 
to be used.
The mass splittings inside the baryon octet and decuplet also get
larger. In most cases this behavior deteriorates the predictions 
inside the baryon octet somewhat, being already too large by a small 
amount in the calculation from the previous section.

These results for the mass spectrum seem to point in the direction 
of a small string tension of about $\sigma=0.08~\text{GeV}^2$  and 
a slight preference for a PV-coupling when the overall agreement
is considered.

\section{Magnetic moments}
\label{sec:magnmom}

Now the influence of the perturbative exchanges on the
mass spectrum has been calculated and the mixing of excited
quark orbitals into the baryon ground state has been estimated
the question arises what might be the consequences for the baryon
magnetics moments. To this objective the expressions 
obtained for the baryon magnetic moments in 
Ref.~\cite{Simonov:2001iz} have to be generalized to arbitrary
quark orbitals to be used within the multichannel calculation.
For the coupling of the meson to the quark we consider two possible
forms, the pseudoscalar (PS) and pseudovector (PV) coupling as
was also done in the calculation of the meson exchanges in
the previous sections.

Following the same procedure as in Ref.~\cite{Simonov:2001iz}
to calculate the major contribution to the baryon magnetic moment
we introduce an external electromagnetic field $\veA$ into the
Hamiltonian equation Eq.~(\ref{eq6}) by minimal substitution,
$\vep_i\rightarrow\vep_i-e_q(i)\veA$, $\veA=\tfrac12(\veH\times\ver)$,
and calculate the energy shift perturbatively,
\be
\Delta E=-\vemu\veH,\qquad\Delta H=-e_q^{(1)}\veal^{(1)} . \veA.
\ee
Thus we find for the magnetic moment operator
\be
\mu_z^{(1)}=-e_q^{(1)}\tfrac{1}{2}\int d^3r_1 
\left\{\phi^*(\ver_1) \left(\vesig^{(1)}\times\ver_1\right)_z\chi(\ver_1)
+\chi^*(\ver_1) \left(\vesig^{(1)}\times\ver_1\right)_z\phi(\ver_1)\right\},
\label{eq24}
\ee
where the single-quark orbital is denoted as
\be 
\psi^{f_1}_{a\alpha}(\ver_1-\ver_0)=
\left(\begin{array}{c}
\phi(\ver_1) \\
\chi(\ver_1)
\end{array}\right).
\ee
The magnetic moment operator Eq.~(\ref{eq24}) can be evaluated by 
rewriting it in terms of spherical harmonics
\be
\frac12\left(\vesig\times\ver\right)=-\frac{1}{2i}\sqrt{\frac{2\pi}{3}}
\left(\sigma_+Y_{1-1}+\sigma_-Y_{11}\right)r,
\ee
after which the angular part can easily be calculated analytically using
Eq.~(\ref{eq51}) which leaves us with a numerical radial integral over $r_1$
(the integrals over $\ver_2$ and $\ver_3$ factorize and drop out),
\begin{multline}
\mu_z=3\mu_z^{(1)}=-3i
\frac1N\int d^3r_1\int d^3r_2 \int d^3r_3\\
{\Psi'_{JM}}^\dagger(\ver_1,\ver_2,\ver_3)
e_q^{(1)}\sqrt{\frac{2\pi}{3}}
\left(\sigma_+^{(1)}Y_{1-1}+\sigma_-^{(1)}Y_{11}\right)r_1
\Psi_{JM}(\ver_1,\ver_2,\ver_3),
\end{multline}
with normalization Eq.~(\ref{eq50}) and symmetrized
baryon wave functions Eq.~(\ref{eq5}).
Because of symmetry considerations we can calculate the contribution
of the first quark only and take the second and third quark into
account by multiplying with a factor of 3.

In the calculation of the baryon mass spectrum we introduced
meson exchanges as an effective interaction representing
the exchange of two quarks. We now study the one-loop effects of
the mesonic fluctuations which give rise to 
modifications of the single-quark current, in particular, to an 
anomalous magnetic moment of the quark.
Near $Q^2=0$ the current can be written as
\be
J_\mu^{\gamma qq}=e_q\gamma_\mu + \kappa_q \frac{ie}{2M_p}\sigma_{\mu\nu}Q^{\nu},
\label{eq25}
\ee
where $\kappa_q = \kappa_s + \kappa_v \tau_z$ for the $u,d$-quark.
From Eq.~(\ref{eq8}) the magnetic moment contribution is found to be
\be
\delta\mu_z^{(1)}=-\int d^3 r_1
\left\{\phi^*(\ver_1)\kappa_q^{(1)}\sigma_z^{(1)}\phi(\ver_1)
-\chi^*(\ver_1)\kappa_q^{(1)}\sigma_z^{(1)}\chi(\ver_1)\right\},
\label{eq26}
\ee
which results in
\be
\delta\mu_z=3\delta\mu_z^{(1)}=3
\frac1N\int d^3r_1\int d^3r_2 \int d^3r_3
{\Psi'_{JM}}^\dagger(\ver_1,\ver_2,\ver_3)
\kappa_q^{(1)}\sigma_z^{(1)}\gamma_0^{(1)}
\Psi_{JM}(\ver_1,\ver_2,\ver_3).
\ee

\begin{figure*}[t!]
\setcaptionmargin{5mm}
\onelinecaptionsfalse
\includegraphics[width=9cm]{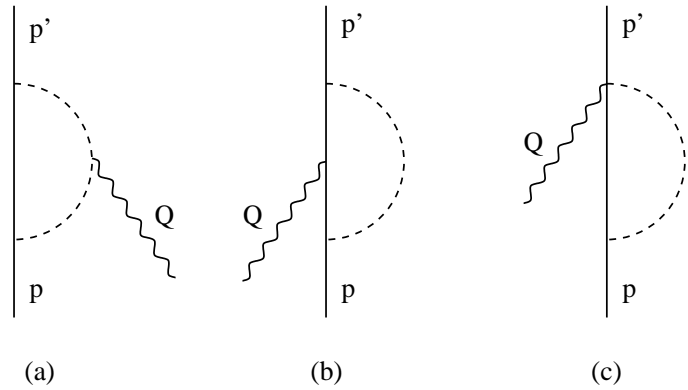}
\captionstyle{normal}
\caption{The diagrams contributing
to the anomalous magnetic moment of 
the single quark. Diagram $c$ is only present when a PV-coupling has been taken. \hfill
\label{fig:anomalpv}}
\end{figure*}

Repeating the procedure followed in Ref.~\cite{Simonov:2001iz} 
we determine the $\kappa$ coefficients in a simple model,
assuming that  the loop corrections are given by only the
one-loop mesonic contributions to the electromagnetic vertex. 
We approximate the single-quark orbital by a free quark 
propagation with a constituent mass given by the ground state 
orbital energy, $\epsilon_0$ shown in Table~\ref{tab:solution}.
With the above simplifying assumptions the calculation amounts
to calculating the magnetic moment contributions of the
diagrams shown in Fig.~\ref{fig:anomalpv}. The diagram with 
the contact interaction is only present when a PV-coupling is
considered.
Assuming a monopole form factor
$F_{\pi qq}(k)=\Lambda_{\pi}^2/\left(\Lambda_{\pi}^2-k^2\right)$ 
we can write,
\begin{subequations}
\label{eq27}
\begin{multline}
J_{\mu}^{(a)}=i\int\frac{d^4 k}{(2\pi)^4}
\Gamma^a_{\pi qq}(k+Q)S_q(p-k)\Delta_{\pi}(k+Q)\Gamma^{ab}_{\gamma\pi\pi,\mu}(k,k+Q)
\displaybreak[0]\\
\times\Delta_{\pi}(k)\Gamma^b_{\pi qq}(-k)
\left(1+
\frac{i}{\Lambda_\pi^2}\frac{F_{\pi qq}\left((k+Q)^2\right)}{\Delta_\pi(k)}+
\frac{i}{\Lambda_\pi^2}\frac{F_{\pi qq}\left((k)^2\right)}{\Delta_\pi(k+Q)}\right),\label{eq27a}
\end{multline}
\be
\shoveleft{J_{\mu}^{(b)}=i\int\frac{d^4 k}{(2\pi)^4}
\Gamma^a_{\pi qq}(k)S_q(p+Q-k)\Gamma_{\gamma qq,\mu}S_q(p-k)\Delta_{\pi}(k)\Gamma^a_{\pi qq}(-k),}\label{eq27b}
\ee
\begin{multline}
J_{\mu}^{(c)}=i\int\frac{d^4 k}{(2\pi)^4}
\biggl(\Gamma^a_{\pi qq}(k)S_q(p+Q-k)\Gamma^a_{\gamma\pi qq,\mu}(-k)+\\
\Gamma^a_{\gamma\pi qq,\mu}(k)S_q(p-k)\Gamma^a_{\gamma\pi qq,\mu}(-k)\biggr)\Delta_\pi(k),
\label{eq27c}
\end{multline}
\end{subequations}
where we use for the propagators
\begin{align}
S_q(p)&=\frac{i\left(\not\! p+M_q\right)}{p^2-M_q^2+i\epsilon},
\label{eq28}\\
\Delta_{\pi}(p)&=\frac{i}{p^2-m_{\pi}^2},
\label{eq29}
\end{align}
and for the vertices
\begin{align}
\Gamma^{ab}_{\gamma\pi\pi,\mu}(k',k)&=
-e\epsilon^{ab3}\left(k_{\mu}+k'_{\mu}\right),
\label{eq30}\\
\Gamma_{\gamma qq,\mu}&=-ie_q\gamma_{\mu},
\label{eq31}\\
\Gamma^a_{\pi qq}(k)&=g_{\pi qq}\gamma_5\tau^aF_{\pi qq}(k),\quad{\rm {(PS)}}.
\label{eq32}
\end{align}
The PV-coupling vertex can be found from Eq.~(\ref{eq32}) by applying 
the replacement Eq.~(\ref{eq16}). In case of a PV-coupling of the meson the 
minimal coupling of the electromagnetic field gives rise to the contact 
interaction
\be
\Gamma^a_{\gamma\pi qq,\mu}(k)=ie\frac{g}{2m_{eff}}\gamma_5\gamma_\mu\tau_b\epsilon_{ba3}F_{\pi qq}(k),\quad{\rm {(PV)}}.
\label{eq33}
\ee
In Eq.~(\ref{eq27a}) an extra term has been added to satisfy the 
Ward-Takahashi identity in second order~\cite{tie}
\be
Q_{\mu}\Gamma^{\mu,(2)}_{\gamma qq}=
e_q\left({S_q^{(2)}}^{-1}(p+Q)-{S_q^{(2)}}^{-1}(p)\right),
\label{eq34}
\ee
where the three point vertex $\Gamma^{\mu,(2)}_{\gamma qq}$ is given
by the sum of the currents Eqs.~(\ref{eq27}).
The currents can now be simplified by shifting the $\gamma_5$'s through the
expression and assuming that the incoming and outgoing quarks are on 
mass-shell. As a result we find
\begin{subequations}
\label{eq35}
\begin{align}
J_{\mu}^{(a,c)} &=-2ie\tau_{z}\left(\gamma^{\nu}C_{\mu\nu}^{(a,c)}+C_{\mu}^{(a,c)}\right),
\label{eq35a}\displaybreak[0]\\
\intertext{and}
J_{\mu}^{(b)}& = ie\frac{1-\tau_z}{2}\left(\gamma^{\nu}C_{\mu\nu}^{(b)}+C_{\mu}^{(b)}\right).
\label{eq35b}
\end{align}
\end{subequations}
From these currents the anomalous magnetic moment has to be extracted.
To isolate this term we first note that 
the tensors $C^{\mu\nu}$ and the vectors $C^{\mu}$ depend only on the 
initial and final momenta.
Therefore they can be written as
\begin{align}
C_{\mu\nu}^{(i)}&=A_1^{(i)}K_{\mu}K_{\nu}+A_2^{(i)}K_{\mu}Q_{\nu}+A_3^{(i)}Q_{\mu}K_{\nu}
+A_4^{(i)}Q_{\mu}Q_{\nu}+A_5^{(i)}g_{\mu\nu},
\label{eq36}\displaybreak[0]\\
C_{\mu}^{(i)}&=B_1^{(i)}K_{\mu}+B_2^{(i)}Q_{\mu}.
\label{eq37}
\end{align}
where $A_n^{(i)}$ and $B_n^{(i)}$ are Lorentz invariants. 
By applying the Gordon decomposition to the current Eq.~(\ref{eq25}) near
$Q^2=0$ it can be seen that the anomalous magnetic moment $\kappa$ is the 
term proportional to $- \frac{e}{2M} K_{\mu}$ with
$K_{\mu}=p_{\mu}+p'_{\mu}$. 
So only the first 
terms, $A_1^{(i)}$ and $B_1^{(i)}$, contribute to the
anomalous magnetic moment. Substituting Eqs.~(\ref{eq36}) and
(\ref{eq37}) in Eq.~(\ref{eq35}) and
taking the initial and final quark on-mass shell we find
for the anomalous magnetic moment corrections
\begin{align}
\kappa^{(a,c)}&=4iM_p\tau_z\left[2m_qA_1^{(a,c)}+B_1^{(a,c)}\right],
\label{eq38}\displaybreak[0]\\
\kappa^{(b)}&=-2iM_p\frac{1-\tau_z}{2}\left[2m_qA_1^{(b)}+B_1^{(b)}\right].
\label{eq39}
\end{align}
Eq.~(\ref{eq38}) corresponds to the
coupling of the photon to the pion and Eq.~(\ref{eq39}) to the 
coupling of the photon to the quark. Formally the contact term
is also represented by Eq.~(\ref{eq38}) in case of a PV-coupling, 
this term however vanishes and does not contribute to the quark 
anomalous magnetic moment.

The Lorentz invariant expressions $A_1^{(i)}$ and $B_1^{(i)}$ can 
immediately be
determined from the tensor $C_{\mu\nu}^{(i)}$. We get
\begin{align}
A_1^{(i)}&=\frac{1}{3K^4}\left(4K^{\mu}K^{\nu}-K^2g^{\mu\nu}\right)C_{\mu\nu}^{(i)},
\label{eq40}\displaybreak[0]\\
\intertext{and}
B_1^{(i)}&=\frac{1}{K^2}K^{\mu}C_{\mu}^{(i)}.
\label{eq41}
\end{align}
Details on the calculation of the integrals and explicit
expressions for $A_1^{(i)}$ and $B_1^{(i)}$
can be found in Appendix \ref{app:anomal}.

The kaon and eta one-loop diagrams can be calculated in similar way. 
The starting point is the expression (\ref{eq27}) again,
where the mass of the pion is replaced by the mass of the kaon and the eta
respectively. In case of the kaon-loop he isospin structure is changed as 
$\tau_z\rightarrow(\tau_z+3 Y)/2$ and 
$(1-\tau_z)/2\rightarrow-(\frac{2}{9}+\frac{4}{3}Y)$ in 
Eqs.~(\ref{eq27}) respectively with $Y$ the hypercharge.
The eta-loop only contributes to the diagram where the photon couples
to the quark as the eta is a charge-neutral meson. Therefore the isospin
structure changes into $\tau_z\rightarrow0$ and 
$(1-\tau_z)/2\rightarrow-\tfrac19+\tfrac16\tau_z+\tfrac12Y$ in 
Eqs.~(\ref{eq27}) respectively.
The coupling constants $g_{K qq}$, $g_{\eta qq}$ and the cutoff masses
$\Lambda_{K}$, $\Lambda_\eta$ are taken as discussed in 
Section~\ref{sec:exchangepotentials} and as shown in 
Table~\ref{tab:parameters}.

From the calculation it is found that the results using a PV-coupling
can easily be related to the outcome using the PS-coupling
\begin{subequations}
\label{eq42}
\begin{align}
A_1^{(a,b)}(PV)&= \left(\frac{m_q+M_q}{2m_{eff}}\right)^2A_1^{(a,b)}(PS),
\\
B_1^{(a,b)}(PV)&= \left(\frac{m_q+M_q}{2m_{eff}}\right)^2B_1^{(a,b)}(PS),
\end{align}
\end{subequations}
where $m_q$ is the constituent mass of the external quark and $M_q$ is 
the constituent mass of the internal quark, both given by the respective 
ground state orbital energy $\epsilon_0$. In case of pionic loops both internal
and external quarks are $u,d$-quarks, $m_q=M_q$. Kaons however change 
$u,d$-quarks into $s$-quarks and back, resulting in $m_q\neq M_q$. 
If the effective mass $m_{eff}$ in the PV-coupling is taken the same 
as the constituent quark mass $m_q$, both couplings 
give the same value. However, from Table~\ref{tab:parameters} it can be seen
that the effective mass differs from the ground state orbital energy $\epsilon_0$
shown in Table~\ref{tab:solution} resulting in different values when
the PS- or PV-coupling is employed. The results are shown in 
Table~\ref{tab:anommagnmom}. The analysis performed shows that the 
contact term does not contribute to the anomalous magnetic moment 
of the quark.

As the contribution from pion exchange currents are predicted to be
small \cite{Simonov:2001iz} we leave them out in first approximation. 

The results on the baryon magnetic moments are shown in 
Tables~\ref{result4channel0}, \ref{result4channel} and 
\ref{result4channelb} for string 
tensions of $\sigma=0.06~\text{GeV}^2$, $\sigma=0.09~\text{GeV}^2$ 
and  $\sigma=0.12~\text{GeV}^2$ respectively. 
The inclusion of the excited quark orbitals decreases the 
baryon magnetic moment.
This behavior results in too low values in case of $\sigma=0.09~\text{GeV}^2$ 
and  $\sigma=0.12~\text{GeV}^2$ while  $\sigma=0.06~\text{GeV}^2$
yields values which are too large.
The best overall agreement is obtained for a string tension in
between, $\sigma=0.08~\text{GeV}^2$.
When a PV-coupling is exploited the anomalous magnetic moment 
contributions are larger which causes an increment of the 
resulting total magnetic moments of the baryons.
Although the results obtained by using either PS- or PV-couplings
are rather similar, 
the PS case seems to produce results slightly closer to experiment.

\section{Conclusions}
\label{sec:conclusion}

In the present paper we have extended the work started in 
Ref.~\cite{Simonov:2001iz} where the field correlator method
was applied to light baryons and magnetic moments were calculated
for the baryon multiplet. The extension comes from the calculation
of the influence of perturbative one-gluon and one-meson exchanges 
on the mass spectrum and magnetic moments of the baryon multiplet.

The described method should be looked at as a second approximation to
calculate both the magnetic moments and the baryon mass spectrum in
the QCD-string model. 
The first approximation is described in Ref.~\cite{Simonov:2001iz}
where no correlations between the quarks were taken into account.
This means that the baryon wave function was described as a product of
single quark orbitals. In the present paper this is partially repaired
by considering one gluon and meson exchanges and taking excited
single-quark orbitals into account. However, effects from neglecting 
the actual position of $\ver_0$ in the Torricelli point and instead 
choosing a fixed value for the parameter $\ver_0$ are not considered
and left for further study.

From the results presented in this paper it appears to be possible
to obtain a reasonable agreement of the baryon magnetic moments in
a region where the predicted masses are close to experiment.
Although there is a small preference for a PS-coupling when the 
magnetic moments are considered, the mass spectrum puts in more
weight in favor of a PV-coupling.
So the best overall agreement is obtained when a PV-coupling
is assumed and a string tension of $\sigma=0.08~\text{GeV}^2$ is
taken.

\begin{acknowledgments}
The work of J.W.\ was supported in part by the Stichting voor
Fundamenteel Onderzoek der Materie (FOM), which is sponsored by
NWO and of J.A.T.\ by the DOE Contract No. DE-AC05-84ER40150
under which SURA operates the Thomas Jefferson National 
Accelerator Facility.
The authors are very grateful to Yu.A.\ Simonov for numerous 
discussions concerning the topic of this paper.
\end{acknowledgments}

\appendix

\section{Symmetric baryon wave functions}
\label{app:sym}

In this Appendix the explicit formulas of the baryon wave function 
are constructed for the nucleon and the $\Delta$ revealing the (iso)spin
structure.
Assuming that all quarks are in their ground state the spin-isospin
wave function has to be symmetric under the exchange of any two quarks.
The color which takes care of the total antisymmetrization is disregarded.

From three (iso)spin-1/2 particles different combinations can be formed which
are denoted as~\cite{Isgur:1978xj,Isgur:1979wd,Capstick:1986bm},
\begin{subequations}
\label{eq43}
\begin{align}
\chi_{3/2}^{S}	&=\left|\tfrac{3}{2}\tfrac{3}{2}\right\rangle	=\uparrow\uparrow\uparrow, 
\label{eq43a}\\
\chi_{1/2}^{S}	&=\left|\tfrac{3}{2}\tfrac{1}{2}\right\rangle	=\sqrt{\frac{1}{3}}\left(\uparrow\uparrow\downarrow+\uparrow\downarrow\uparrow+\downarrow\uparrow\uparrow\right), \label{eq43b}\\
\chi_{1/2}^{\rho}	&=\left|\tfrac{1}{2}\tfrac{1}{2}\right\rangle_{\rho}		=\sqrt{\frac{1}{2}}\left(\downarrow\uparrow\uparrow-\uparrow\downarrow\uparrow\right), \label{eq43c}\\
\chi_{1/2}^{\lambda}	&=\left|\tfrac{1}{2}\tfrac{1}{2}\right\rangle_{\lambda}	=\sqrt{\frac{1}{6}}\left(\uparrow\downarrow\uparrow+\downarrow\uparrow\uparrow-2\uparrow\uparrow\downarrow\right).
\label{eq43d}
\end{align}
\end{subequations}
Negative spin functions are obtained by flipping all spins. 
It can be seen that Eqs.~(\ref{eq43a}), (\ref{eq43b}) and (\ref{eq43d})
are obtained by coupling a $j_{12}=1$ state with $j_3=\tfrac12$ whereas
Eq.~(\ref{eq43c}) is formed from a $j_{12}=0$ state and a $j_3=\tfrac12$ 
state.
Isospin functions can
similarly be written down resulting in $\eta_{m_{i}}^{S}$, 
$\eta_{m_{i}}^{\rho}$, and $\eta_{m_{i}}^{\lambda}$. Eqs.~(\ref{eq43}) 
contain all possible combinations and are orthonormal.
The states $S$ are total symmetric while  $\rho$ and $\lambda$ are 
mixed-symmetric states. When the interchange of particle
$i$ and $j$ is denoted by $P_{ij}$ they behave as
\begin{align}
P_{12}\chi^{S}&=\chi^{S}, &  P_{13}\chi^{S}&=\chi^{S} \\
P_{12}\chi^{\rho}&=-\chi^{\rho},      & P_{13}\chi^{\rho}&=-\frac{1}{2}\sqrt{3}\chi^{\lambda}+\frac{1}{2}\chi^{\rho}, \\
P_{12}\chi^{\lambda}&=\chi^{\lambda}, & P_{13}\chi^{\lambda}&=-\frac{1}{2}\sqrt{3}\chi^{\rho}-\frac{1}{2}\chi^{\lambda}.
\end{align}
The states  $\rho$ and $\lambda$ are clearly not symmetric under the
permutation of any two quarks. However some specific combination of 
$\chi^{\rho,\lambda}$ and $\eta^{\rho,\lambda}$ is symmetric, actually,
from the states Eqs.~(\ref{eq43}) only two totally symmetric states can be 
formed,
\begin{align}
\Delta(m_{s},m_{i})&=\chi^{S}_{m_{s}}\cdot\eta^{S}_{m_{i}},& J&=3/2, & I&=3/2, \label{eq44}\\
N(m_{s},m_{i})&=\frac{1}{\sqrt{2}}\left(\chi^{\rho}_{m_{s}}\eta^{\rho}_{m_{i}}+\chi^{\lambda}_{m_{s}}\eta^{\lambda}_{m_{i}}\right), & J&=1/2, & I&=1/2.\label{eq45}
\end{align}
Eq.~(\ref{eq44}) and Eq.~(\ref{eq45}) represent the $\Delta$ and the
nucleon respectively.
The formalism can be extended to the total baryon octet and decuplet by 
including the $s$-quark in writing down a complete orthonormal set.

\section{Calculation of the exchange potentials}
\label{app:exchpot}

In the calculation of the matrix elements, Eqs.~(\ref{eq13}, 
\ref{eq24}, \ref{eq26}), use 
has been made of the partial wave decomposition of the single quark 
orbitals and the expansion of the operator in spherical 
harmonics. This enables an easy analytic calculation of the angles.

The single quark orbital is decomposed as~\cite{simtjo2}
\be
\psi_{a\alpha}^{f}(\ver)=
\left|njlm_j\right>\otimes\left|im_i\right>=
\left(\begin{array}{c}
g_n(r)\Omega_{jlm_j} \\ if_n(r)\Omega_{j\tilde{l}m_j}
\end{array}\right)
\otimes\eta_{m_i},\qquad \tilde{l}=2j-l,
\ee
with
\be
\Omega_{jlm_j}=\sum_{m_s}C(l\tfrac12j;m_j-m_s,m_s,m_j)Y_{lm_j-m_s}\chi_{m_s}
\ee
where $\chi_{m_s}$ ($\eta_{m_i}$) is the (iso)spin-function and $C$ 
is the Clebsch-Gordon coefficient in the notation of Rose~\cite{rose}. 

A total symmetric baryon wave function can be composed out of the
single quark orbitals as described in section~\ref{sec:mccalculation}.
This procedure is summarized as
\be
\Psi_{JM}(\ver_1,\ver_2,\ver_3) = \Gamma^{\alpha\beta\gamma}_{JM} (f_1f_2f_3) 
 \psi^{f_1}_{\alpha}(\ver_1) 
 \psi^{f_2}_{\beta}(\ver_2) 
 \psi^{f_3}_{\gamma}(\ver_3),
%\Psi_{J,I}=\sum_{\text{all indices}}\Gamma^{n_1n_2n_3}_{j_1j_2j_3,i_1i_2i_3}(J,I)\prod_{k=1}^{3}
%\left|n_kj_kl_km_{j_k}\right>\otimes\left|i_km_{i_k}\right>,
\ee
where the $\Gamma$ takes care of the symmetrization, $f_i$ are the
flavor indices, $\alpha$, $\beta$ and $\gamma$ indicate the quark excitation
and the color indices are left out for simplicity. It is understood that the 
baryon wave function is in a color singlet state.

The energy shift can quite easily be calculated 
after some modifications of the exchange potential.
Let us consider the equations written in coordinate space, Eqs.~(\ref{eq15}).
Then, expand the potentials in terms of spherical harmonics
$Y_{lm}$ as
\begin{align}
V(\ver_2,\ver_3) & =  \sum_{l=0}^{\infty} V_l(r_2,r_3)\frac{2l+1}{4\pi}P_l(x) 
\notag\\
& =  \sum_{l=0}^{\infty} V_l(r_2,r_3) \sum_{m=-l}^l Y_{lm}^*(\Omega_2)Y_{lm}(\Omega_3)
\label{eq46}
\end{align}
with $P_l$ the Legendre polynomials and $x=\cos(\gamma)$ the angle between the
vectors $\ver_2$ and $\ver_3$. 
The function $V_l(r_2,r_3)$ can be found by using the orthonormality condition
of the Legendre polynomials
\be
V_l(r_2,r_3)=\frac{4\pi}{2}\int_{-1}^{1}V(\ver_2,\ver_3)P_l(x)dx
\label{eq47}
\ee
In the special case of the Coulomb potential 
$V(\ver_2,\ver_3)=1/\left|\ver_2-\ver_3\right|$ the integral
can be done analytically and the expansion looks like
\be
V_l(r_2,r_3)=\frac{4\pi}{2l+1}\frac{r_{<}^l}{r_{>}^{l+1}},
\label{eq48}
\ee
with $r{<}$ ($r{>}$) the smaller (greater) of $r_2$ and $r_3$.
The advantage of this expansion is the easy analytic
evaluation of the integrals over the angles appearing in the calculation of
the matrix elements in Eq.~(\ref{eq13}).

The matrix element Eq.~(\ref{eq13}) can now be written as
\begin{multline}
\label{eq49}
\left\langle\Psi'_{JM}\left|H_{\rm int}^{(23)}\right|\Psi_{JM}\right\rangle=
\frac1N\int d^3r_1\int d^3r_2 \int d^3r_3
{\Psi'_{JM}}^\dagger(\ver_1,\ver_2,\ver_3)\\
\times\left(\Lambda^+(1)\Lambda^+(2)\Lambda^+(3)+\Lambda^-(1)\Lambda^-(2)\Lambda^-(3)
\right)
\gamma_0(2)\gamma_0(3)V(\ver_2,\ver_3)
\Psi_{JM}(\ver_1,\ver_2,\ver_3)
\end{multline}
with the normalization
\be
N=
\int d^3r_1\int d^3r_2 \int d^3r_3
{\Psi'_{JM}}^\dagger(\ver_1,\ver_2,\ver_3)
\Psi_{JM}(\ver_1,\ver_2,\ver_3)
\label{eq50}
\ee
and the projection matrices $\Lambda^\pm$ defined as Eq.~(\ref{eq11}).

The integral over the first quark factorizes and drops out. The remaining
part contains two angular integrals, $d\Omega_2$ and $d\Omega_3$, over 
three spherical harmonics each.
The product of three spherical harmonics can analytically be evaluated
as (see for example Ref.~\cite{rose})
\be
\int d\Omega Y^*_{l'm'}Y_{LM}Y_{lm}=\sqrt{\frac{\left(2l+1\right)\left(2L+1\right)}{4\pi\left(2l'+1\right)}}C(lLl';mMm')C(lLl';000).
\label{eq51}
\ee
The remaining radial integral over $r_2$ and $r_3$ is done numerically.

In case of PV-coupling extra matrices $\not\! k$ in Eqs.~(\ref{eq16}) 
are added to Eq.~(\ref{eq49}) and become derivatives in 
coordinate space.
As the derivatives act on the wave functions the actual potential 
$V_{ope}^{(23)}(\ver_2-\ver_3)$ can still be expanded in terms of 
spherical harmonics in the exactly same way as is done for PS-coupling,
which results in
\begin{multline}
\label{eq52}
\left\langle\Psi'_{JM}\left|H_{\rm int}^{(23)}\right|\Psi_{JM}\right\rangle=
\frac1N\int d^3r_1\int d^3r_2 \int d^3r_3
{\Psi'_{JM}}^\dagger(\ver_1,\ver_2,\ver_3)\\
\shoveright{
\times\left(\Lambda^+(1)\Lambda^+(2)\Lambda^+(3)+\Lambda^-(1)\Lambda^-(2)\Lambda^-(3)
\right)
\gamma_0(2)\gamma_0(3)
}
\\
\shoveleft{
\times
\left(\begin{array}{cc}
i\vesig^{(2)}\leftnab_{2} & \varepsilon_{n'_2}^{(2)} \\
\varepsilon_{n'_2}^{(2)} & i\vesig^{(2)}\leftnab_{2} \\
\end{array}\right)_{(2)}
\left(\begin{array}{cc}
-i\vesig^{(3)}\leftnab_{3} & -\varepsilon_{n'_3}^{(3)} \\
-\varepsilon_{n'_3}^{(3)} & -i\vesig^{(3)}\leftnab_{3} \\
\end{array}\right)_{(3)}
V(\ver_2,\ver_3)
}
\\
\times
\left(\begin{array}{cc}
i\vesig^{(2)}\rightnab_{2} & -\varepsilon_{n_2}^{(2)} \\
-\varepsilon_{n_2}^{(2)} & i\vesig^{(2)}\rightnab_{2} \\
\end{array}\right)_{(2)}
\left(\begin{array}{cc}
-i\vesig^{(3)}\rightnab_{3} & \varepsilon_{n_3}^{(3)} \\
\varepsilon_{n_3}^{(3)} & -i\vesig^{(3)}\rightnab_{3} \\
\end{array}\right)_{(3)}
\Psi_{JM}(\ver_1,\ver_2,\ver_3)
\end{multline}
The arrows point in the direction in which the derivatives operate.
Again the integral over $\ver_1$ factorizes and drops out.
The derivative on the wave function has to be
calculated, which can be done by using
\be
\vesig\venab g(r)\Omega_{jlm}=-\left(\frac{d}{dr}g(r)+\frac{1+\kappa}{r}g(r)\right)\Omega_{j\tilde{l}m}
\ee
with $\tilde{l}=2j-l$, $\Omega_{j\tilde{l}m}=-(\vesig\hat{r})\Omega_{jlm}$ 
and $\kappa=\pm(j+\tfrac12)$ as $j=l\pm\tfrac12$.
The integral over the angles can again be evaluated using Eq.~(\ref{eq51}),
while the integral over the radial wave functions, containing also derivatives
on the radial wave functions, is done numerically.

\section{Anomalous magnetic moment contributions from meson loops}
\label{app:anomal}

In this appendix explicit formulas on the contribution 
to the anomalous magnetic moment of one-loop diagrams are shown.
In the first subsection the pseudoscalar coupling has been used, 
Eq.~(\ref{eq32}),
in the second subsection the pseudovector coupling.
In the last subsection 
some useful formulas on the calculation of the integrals are
given.

\subsection{Pseudoscalar coupling}

Our starting point is the electromagnetic currents, corresponding to the
one-loop diagrams shown in Fig.~\ref{fig:anomalpv}, assuming a $\gamma_5$
coupling of the pion to the quark,
\begin{align}
J_{\mu}^{(a)} &= -2ig_{\pi qq}^{2}e\tau_z
\int 
\frac{d^4k}{\left(2\pi\right)^4}
\frac{\gamma_5\left(\not\! p-\not\! k+M_q\right)\gamma_5\left(2k_{\mu}+Q_{\mu}\right)}
{\left[(p-k)^2-M_q^2+i\epsilon\right]\left[k^2-m_{\pi}^2+i\epsilon\right]\left[\left(k+Q\right)^2-m_{\pi}^2+i\epsilon\right]}
\nonumber
\\
&\quad\times\frac{\Lambda_{\pi}^2}{k^2-\Lambda_{\pi}^2}
\frac{\Lambda_{\pi}^2}{\left(k+Q\right)^2-\Lambda_{\pi}^2}
\left(1+\frac{k^2-m_{\pi}^2}{\left(k+Q\right)^2-\Lambda_{\pi}^2}+
\frac{\left(k+Q\right)^2-m_{\pi}^2}{k^2-\Lambda_{\pi}^2}\right)
\nonumber\displaybreak[0]\\
& \equiv -2ie\tau_{z}\left(\gamma^{\nu}C_{\mu\nu}^{(a)}+C_{\mu}^{(a)}\right),
\label{eq54}\displaybreak[0]\\
\intertext{and}
J_{\mu}^{(b)}&=-ig_{\pi qq}^{2}e\frac{1-\tau_z}{2}
\int
\frac{d^4k}{\left(2\pi\right)^4}.
\frac{\gamma_5\left(\not\! p'-\not\! k+M_q\right)\gamma_{\mu}\left(\not\! p-\not\! k+M_q\right)\gamma_5}
{\left[(p'-k)^2-M_q^2+i\epsilon\right]\left[(p-k)^2-M_q^2+i\epsilon\right]\left[k^2-m_{\pi}^2+i\epsilon\right]}
\nonumber\\
&\quad\times\left(\frac{\Lambda_{\pi}^2}{k^2-\Lambda_{\pi}^2}\right)^2
\nonumber\displaybreak[0]\\
& \equiv  ie\frac{1-\tau_z}{2}\left(\gamma^{\nu}C_{\mu\nu}^{(b)}+C_{\mu}^{(b)}\right).
\label{eq55}
\end{align}
In writing these equations use has been made 
of explicit evaluation of the $\gamma$-matrix algebra and
of the approximation that the initial and final quark are on-mass shell.
To be able to discuss more general diagrams the masses of the external quark $m_q$ 
and the intermediate quark $M_q$ are taken differently. In case of the pionic
fluctuations of the $u,d$-quark the equations can be reduced using $M_q=m_q$.
Since we have assumed a finite form factor at the $\pi q q$ vertex, 
similar as in the two-body current case, the two additional terms 
are needed
in the last factor of Eq.~(\ref{eq54}) to satisfy the Ward-Takahashi identity,
Eq.~(\ref{eq34}).
From these currents the anomalous magnetic moment has to be extracted.
As was discussed in Section~\ref{sec:magnmom} this can be done
by calculation the Lorentz-invariant terms  $A_1^{(i)}$ and $B_1^{(i)}$
which are found as described in Eqs.~(\ref{eq40}) and (\ref{eq41}).
The expressions for $A_1^{(a)}$ and $B_1^{(a)}$ are
\begin{align}
A_1^{(a)}&=\frac{g_{\pi qq}^2}{6 m_q^4}\int\frac{4\left(p\cdot k\right)^2-p^2k^2}{\left[k^2-2pk+m_q^2-M_q^2+i\epsilon\right]\left[k^2-m_{\pi}^2+i\epsilon\right]^2}\nonumber\\
&\quad\times\left(\frac{\Lambda_{\pi}^2}{\Lambda_{\pi}^2-k^2}\right)^2\left(1+2\frac{k^2-m_{\pi}^2}{k^2-\Lambda_{\pi}^2}\right)\frac{d^4k}{\left(2\pi\right)^4}
\displaybreak[0]\\
&=\frac{-i}{32\pi^2}g_{\pi qq}^2\int_0^1d\alpha \alpha\left(1-\alpha\right)^2
\left[\left(\frac{1}{F_{m_{\pi}}}-\frac{1}{F_{\Lambda_{\pi}}}\right)\left(\frac{\Lambda_{\pi}^2}{\Lambda_{\pi}^2-m_{\pi}^2}\right)^2-\alpha\frac{\Lambda_{\pi}^2}{F_{\Lambda_{\pi}}^2}\frac{\Lambda_{\pi}^2}{\Lambda_{\pi}^2-m_{\pi}^2}\right],
\displaybreak[0]\\
\intertext{and}
B_1^{(a)}&=\frac{g_{\pi qq}^2}{m_q^2}\int\frac{\left(M_q-m_q\right)p\cdot k}{\left[k^2-2pk+m_q^2-M_q^2+i\epsilon\right]\left[k^2-m_{\pi}^2+i\epsilon\right]^2}\nonumber\\
&\quad\times\left(\frac{\Lambda_{\pi}^2}{\Lambda_{\pi}^2-k^2}\right)^2\left(1+2\frac{k^2-m_{\pi}^2}{k^2-\Lambda_{\pi}^2}\right)\frac{d^4k}{\left(2\pi\right)^4}
\displaybreak[0]\\
&=\frac{-i}{16\pi^2}g_{\pi qq}^2\left(M_q-m_q\right)\int_0^1d\alpha \alpha\left(1-\alpha\right)\nonumber\\
&\quad\times\left[\left(\frac{1}{F_{m_{\pi}}}-\frac{1}{F_{\Lambda_{\pi}}}\right)\left(\frac{\Lambda_{\pi}^2}{\Lambda_{\pi}^2-m_{\pi}^2}\right)^2-\alpha\frac{\Lambda_{\pi}^2}{F_{\Lambda_{\pi}}^2}\frac{\Lambda_{\pi}^2}{\Lambda_{\pi}^2-m_{\pi}^2}\right].
\end{align}
The expressions for $A_1^{(b)}$ and $B_1^{(b)}$ are
\begin{align}
A_1^{(b)}&=\frac{g_{\pi qq}^2}{6 m_q^4}\int\frac{4\left(p\cdot k\right)^2-p^2k^2}{\left[k^2-2pk+m_q^2-M_q^2+i\epsilon\right]^2\left[k^2-m_{\pi}^2+i\epsilon\right]}
\left(\frac{\Lambda_{\pi}^2}{\Lambda_{\pi}^2-k^2}\right)^2\frac{d^4k}{\left(2\pi\right)^4}
\displaybreak[0]\\
&=\frac{-i}{32\pi^2}g_{\pi qq}^2\int_0^1d\alpha \left(1-\alpha\right)^3
\left[\left(\frac{1}{F_{m_{\pi}}}-\frac{1}{F_{\Lambda_{\pi}}}\right)\left(\frac{\Lambda_{\pi}^2}{\Lambda_{\pi}^2-m_{\pi}^2}\right)^2-\alpha\frac{\Lambda_{\pi}^2}{F_{\Lambda_{\pi}}^2}\frac{\Lambda_{\pi}^2}{\Lambda_{\pi}^2-m_{\pi}^2}\right],
\displaybreak[0]\\
\intertext{and}
B_1^{(b)}&=\frac{g_{\pi qq}^2}{m_q^2}\int\frac{\left(M_q-m_q\right)p\cdot k}{\left[k^2-2pk+m_q^2-M_q^2+i\epsilon\right]^2\left[k^2-m_{\pi}^2+i\epsilon\right]}
\left(\frac{\Lambda_{\pi}^2}{\Lambda_{\pi}^2-k^2}\right)^2\frac{d^4k}{\left(2\pi\right)^4}
\displaybreak[0]\\
&=\frac{-i}{16\pi^2}g_{\pi qq}^2\left(M_q-m_q\right)\int_0^1d\alpha \left(1-\alpha\right)^2\nonumber\\
&\quad\times\left[\left(\frac{1}{F_{m_{\pi}}}-\frac{1}{F_{\Lambda_{\pi}}}\right)\left(\frac{\Lambda_{\pi}^2}{\Lambda_{\pi}^2-m_{\pi}^2}\right)^2-\alpha\frac{\Lambda_{\pi}^2}{F_{\Lambda_{\pi}}^2}\frac{\Lambda_{\pi}^2}{\Lambda_{\pi}^2-m_{\pi}^2}\right].
\end{align}
In these formulas $F_m$ is defined as
\be
F_m=\left(1-\alpha\right)^2 m_q^2+\left(1-\alpha\right)\left(M_q^2-m_q^2\right)+\alpha m^2.
\label{eq56}
\ee
Frequent use has been made of the formulas listed in 
subsection~\ref{sec:1loopintegrals}.

\subsection{Pseudovector coupling}

In case of pseudovector coupling the PS-vertex has to be changed into 
the PV-vertex by applying Eq.~(\ref{eq16}).
The currents can be reduced to
\begin{align}
\label{eq57}
J_{\mu}^{(a)} & =  -2i\frac{g_{\pi qq}^{2}}{4m_{eff}^2}e\tau_{z}\int 
\frac{d^4k}{\left(2\pi\right)^4}
\left\{\frac{2k_{\mu}\not\! k\left(k^2-2p\cdot k+2m_q\left(m_q+M_q\right)\right)}
{\left[k^2-2pk+m_q^2-M_q^2+i\epsilon\right]\left[k^2-m_{\pi}^2+i\epsilon\right]^2}\right.\nonumber\\
&\quad\left.+\frac{2k_{\mu}\left(k^2-2p\cdot k\right)\left(M_q+m_q\right)}
{\left[k^2-2pk+m_q^2-M_q^2+i\epsilon\right]\left[k^2-m_{\pi}^2+i\epsilon\right]^2}\right\}
\left(\frac{\Lambda_{\pi}^2}{\Lambda_{\pi}^2-k^2}\right)^2 
\left(1+2\frac{k^2-m_{\pi}^2}{k^2-\Lambda_{\pi}^2}\right)
\nonumber\displaybreak[0]\\
& \equiv  -2ie\tau_{z}\left(\gamma^{\nu}C_{\mu\nu}^{(a)}+C_{\mu}^{(a)}\right),
\displaybreak[0]\\
\intertext{and}
J_{\mu}^{(b)} & =  ie\frac{1-\tau_z}{2}\frac{g_{\pi qq}^{2}}{4m_{eff}^2}\int 
\frac{d^4k}{\left(2\pi\right)^4}
\frac{\not\! k\left(m_q+M_q\right)+k^2-2p'\cdot k}{\left[k^2-2p'\cdot k+m_q^2-M_q^2+i\epsilon\right]}\gamma_{\mu}\nonumber\\
&\quad\times\frac{\not\! k\left(m_q+M_q\right)+k^2-2p\cdot k}{\left[k^2-2p\cdot k+m_q^2-M_q^2+i\epsilon\right]}
\frac{1}{\left[k^2-m_{\pi}^2+i\epsilon\right]}\left(\frac{\Lambda_{\pi}^2}{\Lambda_{\pi}^2-k^2}\right)^2 
\nonumber\displaybreak[0]\\
& \equiv   ie\frac{1-\tau_z}{2}\left(\gamma^{\nu}C_{\mu\nu}^{(b)}+C_{\mu}^{(b)}\right),
\displaybreak[0]\\
\intertext{and}
J_{\mu}^{(c)} & =  2i\frac{g_{\pi qq}^{2}}{4m_{eff}^2}e\tau_{z}\int 
\frac{d^4k}{\left(2\pi\right)^4}\left(
\frac{\left(m_q+M_q\right)\not\! k+k^2-2p'\cdot k}{\left[k^2-2p'\cdot k+m_q^2-M_q^2+i\epsilon\right]}\gamma_{\mu}\right.\nonumber\\
&\quad\left.+\gamma_{\mu}\frac{\left(m_q+M_q\right)\not\! k+k^2-2p\cdot k}{\left[k^2-2p\cdot k+m_q^2-M_q^2+i\epsilon\right]}\right)
\frac{1}{\left[k^2-m_{\pi}^2+i\epsilon\right]}\left(\frac{\Lambda_{\pi}^2}{\Lambda_{\pi}^2-k^2}\right)^2 
\displaybreak[0]\\
& \equiv -2ie\tau_z\left(\gamma^{\nu}C_{\mu\nu}^{(c)}+C_{\mu}^{(c)}\right).
\end{align}
Again the Ward-Takahashi identity requires the extra term in Eq.~(\ref{eq57}).
The expressions for $A_1^{(a)}$ and $B_1^{(a)}$ are
\begin{align}
A_1^{(a)}&=\frac{g_{\pi qq}^2}{24 m_{eff}^2m_q^4}\int\frac{\left(4\left(p\cdot k\right)^2-p^2k^2\right)\left(k^2-2p\cdot k+2m_q\left(m_q+M_q\right)\right)}{\left[k^2-2pk+m_q^2-M_q^2+i\epsilon\right]\left[k^2-m_{\pi}^2+i\epsilon\right]^2}\nonumber\\
&\quad\times\left(\frac{\Lambda_{\pi}^2}{\Lambda_{\pi}^2-k^2}\right)^2\left(1+2\frac{k^2-m_{\pi}^2}{k^2-\Lambda_{\pi}^2}\right)\frac{d^4k}{\left(2\pi\right)^4}
\displaybreak[0]\\
&=\frac{-i}{32\pi^2}\frac{g_{\pi qq}^2}{4m_{eff}^2}\left(M_q+m_q\right)^2\int_0^1d\alpha \alpha\left(1-\alpha\right)^2\nonumber\\
&\quad\times\left[\left(\frac{1}{F_{m_{\pi}}}-\frac{1}{F_{\Lambda_{\pi}}}\right)\left(\frac{\Lambda_{\pi}^2}{\Lambda_{\pi}^2-m_{\pi}^2}\right)^2-\alpha\frac{\Lambda_{\pi}^2}{F_{\Lambda_{\pi}}^2}\frac{\Lambda_{\pi}^2}{\Lambda_{\pi}^2-m_{\pi}^2}\right],
\displaybreak[0]\\
\intertext{and}
B_1^{(a)}&=\frac{g_{\pi qq}^2}{4 m_{eff}^2m_q^2}\int\frac{\left(M_q+m_q\right)\left(k^2-2p\cdot k\right)p\cdot k}{\left[k^2-2pk+m_q^2-M_q^2+i\epsilon\right]\left[k^2-m_{\pi}^2+i\epsilon\right]^2}\nonumber\\
&\quad\times\left(\frac{\Lambda_{\pi}^2}{\Lambda_{\pi}^2-k^2}\right)^2\left(1+2\frac{k^2-m_{\pi}^2}{k^2-\Lambda_{\pi}^2}\right)\frac{d^4k}{\left(2\pi\right)^4}
\displaybreak[0]\\
&=\frac{-i}{16\pi^2}\frac{g_{\pi qq}^2}{4m_{eff}^2}\left(M_q-m_q\right)\left(m_q+M_q\right)^2\int_0^1d\alpha \alpha\left(1-\alpha\right)\nonumber\\
&\quad\times\left[\left(\frac{1}{F_{m_{\pi}}}-\frac{1}{F_{\Lambda_{\pi}}}\right)\left(\frac{\Lambda_{\pi}^2}{\Lambda_{\pi}^2-m_{\pi}^2}\right)^2-\alpha\frac{\Lambda_{\pi}^2}{F^2_{\Lambda_{\pi}}}\frac{\Lambda_{\pi}^2}{\Lambda_{\pi}^2-m_{\pi}^2}\right].
\end{align}
The expressions for $A_1^{(b)}$ and $B_1^{(b)}$ are
\begin{align}
A_1^{(b)}&=\frac{g_{\pi qq}^2}{24 m_{eff}^2m_q^4}\int\frac{\left(4\left(p\cdot k\right)^2-p^2k^2\right)\left(m_q+M_q\right)^2}{\left[k^2-2pk+m_q^2-M_q^2+i\epsilon\right]^2\left[k^2-m_{\pi}^2+i\epsilon\right]}
\left(\frac{\Lambda_{\pi}^2}{\Lambda_{\pi}^2-k^2}\right)^2\frac{d^4k}{\left(2\pi\right)^4}
\displaybreak[0]\\
&=\frac{-i}{32\pi^2}\frac{g_{\pi qq}^2}{4m_{eff}^2}\left(m_q+M_q\right)^2\int_0^1d\alpha \left(1-\alpha\right)^3\nonumber\\
&\quad\times\left[\left(\frac{1}{F_{m_{\pi}}}-\frac{1}{F_{\Lambda_{\pi}}}\right)\left(\frac{\Lambda_{\pi}^2}{\Lambda_{\pi}^2-m_{\pi}^2}\right)^2-\alpha\frac{\Lambda_{\pi}^2}{F_{\Lambda_{\pi}}^2}\frac{\Lambda_{\pi}^2}{\Lambda_{\pi}^2-m_{\pi}^2}\right],
\displaybreak[0]\\
\intertext{and}
B_1^{(b)}&=\frac{g_{\pi qq}^2}{4m_{eff}^2m_q^2}\int\frac{\left(M_q+m_q\right)p\cdot k\left(k^2-2p\cdot k\right)}{\left[k^2-2pk+m_q^2-M_q^2+i\epsilon\right]^2\left[k^2-m_{\pi}^2+i\epsilon\right]}
\left(\frac{\Lambda_{\pi}^2}{\Lambda_{\pi}^2-k^2}\right)^2\frac{d^4k}{\left(2\pi\right)^4}
\displaybreak[0]\\
&=\frac{-i}{16\pi^2}\frac{g_{\pi qq}^2}{4m_{eff}^2}\left(M_q-m_q\right)\left(m_q+M_q\right)^2\int_0^1d\alpha \left(1-\alpha\right)^2\nonumber\\
&\quad\times\left[\left(\frac{1}{F_{m_{\pi}}}-\frac{1}{F_{\Lambda_{\pi}}}\right)\left(\frac{\Lambda_{\pi}^2}{\Lambda_{\pi}^2-m_{\pi}^2}\right)^2-\alpha\frac{\Lambda_{\pi}^2}{F_{\Lambda_{\pi}}^2}\frac{\Lambda_{\pi}^2}{\Lambda_{\pi}^2-m_{\pi}^2}\right].
\end{align}
The current resulting from the contact term does not contribute to 
the anomalous magnetic moment, $A_1^{(c)}=0$ and
$ B_1^{(c)}=0$.
In the formulas above $F_m$ is defined as before (Eq.~(\ref{eq56})).
Although the expressions for the currents become much more involved
using a PV-coupling, the final expression can simply be written in 
terms of the result from the previous calculation as shown in 
Eqs.~(\ref{eq42}).

\subsection{Useful formulas
\label{sec:1loopintegrals}}

In the calculation of one-loop integrals frequent use has been 
made of the Feynman parameterization
\be
\frac{1}{ab}=\int_0^1d\alpha\frac{1}{\left[\alpha a+(1-\alpha)b\right]^2}.
\ee
This formula can be generalized to
\begin{multline}
\frac{1}{a_1a_2\ldots a_n}=
(n-1)!\int_0^1d\alpha_1\int_0^{1-\alpha_1}d\alpha_2\ldots\int_0^{1-\alpha_1-\alpha_2\ldots-\alpha_{n-2}}d\alpha_{n-1}\\
\times\frac{1}{\left[\alpha_1 a_1+\alpha_2 a_2+\ldots\left(1-\alpha_1-\alpha_2\ldots-\alpha_{n-1}\right) a_n\right]^n}.
\end{multline}
Which can be proven by induction.
All loop integrals in the text can be reduced to one of the following forms~\cite{Tjon:2000sa}
\be
\int\frac{d^4k}{(2\pi)^4}\left(\frac{1}{\left[(k-l)^2-F_m\right]^2}-\frac{1}{\left[(k-l)^2-F_{\Lambda_{\pi}}\right]^2}\right)\left\{1,k^{\mu}\right\}=
\frac{-i}{16\pi^2}\ln\left(\frac{F_m}{F_{\Lambda_{\pi}}}\right)\left\{1,l^{\mu}\right\},
\ee
\begin{multline}
\int\frac{d^4k}{(2\pi)^4}\left(\frac{1}{\left[(k-l)^2-F_m\right]^3}-\frac{1}{\left[(k-l)^2-F_{\Lambda_{\pi}}\right]^3}\right)k^{\mu}k^{\nu}=\\
\frac{-i}{32\pi^2}\left(-\frac12\ln\left(\frac{F_m}{F_{\Lambda_{\pi}}}\right)g^{\mu\nu}+\left(\frac{1}{F_m}-\frac{1}{F_{\Lambda_{\pi}}}\right)l^{\mu}l^{\nu}\right).
\end{multline}
If the previous formulas do not apply, one can use $(n\geq 3)$
\begin{multline}
\int\frac{d^4k}{(2\pi)^4}\frac{1}{\left[(k-l)^2-F+i\eta\right]^n}\left\{1,k^{\mu},k^{\mu}k^{\nu}\right\}=\\
\frac{i(-1)^n}{(n-1)(n-2)16\pi^2}\left\{\frac{1}{F^{n-2}},\frac{l^{\mu}}{F^{n-2}},\frac{l^{\mu}l^{\nu}}{F^{n-2}}-\frac{g^{\mu\nu}}{2(n-3)F^{n-3}}\right\}.
\end{multline}

\newpage

%\bibliographystyle{maik}
%\bibliography{rever}

\newpage
%
% Figures
%

%%++ Figure:1

\newpage

%%++ Figure:2

\newpage
%
% Tables
%

%%++ Table:1
\begin{table}
\setcaptionmargin{0mm}
\onelinecaptionsfalse
\captionstyle{flushleft}
\caption{
Ground state energy $\epsilon_0$ in units of MeV of the single-particle 
orbitals (spo) for various values of $\sigma$ for both the $u,d$-quark and 
the $s$-quark. For the current masses the values $m_u=m_d=5~MeV$ and 
$m_s=200~MeV$ are adopted.
\label{tab:solution}}
\bigskip
\begin{tabular}{|c|r|r|r|r|r|r|}
\hline
spo & \multicolumn{2}{c|}{$\sigma=0.06~\text{GeV}^2$} & \multicolumn{2}{c|}{$\sigma=0.09~\text{GeV}^2$} & \multicolumn{2}{c|}{$\sigma=0.12~\text{GeV}^2$}\\
\cline{2-7}
&  $\epsilon_0(u,d)$  &  $\epsilon_0(s)$  &  $\epsilon_0(u,d)$ &  $\epsilon_0(s)$  &  $\epsilon_0(u,d)$  &  $\epsilon_0(s)$ \\
\hline\hline
$0(1/2)^{++}$ & 243 & 388 & 297 & 440 & 342 & 482 \\
$1(1/2)^{++}$ & 435 & 560 & 535 & 656 & 617 & 736 \\
$2(1/2)^{++}$ & 572 & 692 & 704 & 821 & 813 & 928 \\
$0(1/2)^{-+}$ & 375 & 497 & 460 & 579 & 531 & 646 \\
$0(3/2)^{++}$ & 339 & 479 & 415 & 553 & 478 & 614 \\
$1(3/2)^{++}$ & 501 & 626 & 616 & 739 & 711 & 832 \\
$0(3/2)^{-+}$ & 453 & 576 & 558 & 677 & 644 & 761 \\
$0(5/2)^{++}$ & 419 & 556 & 514 & 650 & 593 & 728 \\
\hline
\end{tabular}
\end{table}

\newpage

%%++ Table:2
\begin{table}
\setcaptionmargin{0mm}
\onelinecaptionsfalse
\captionstyle{flushleft}
\caption{ The parameters involved in the meson quark interaction
used for the calculation with the axial coupling constant $g_A^q=1$. 
The decay constants $f_m$ and all (cutoff) masses are in units of MeV, the
coupling constants $f_{mqq}$ and $g_{mqq}$ are dimensionless.
\label{tab:parameters}}
\bigskip
\begin{tabular}{|c|r|r|d|r|r|d|r|d|r|d|}
\hline
m & $M_m$ & $f_m$ & \multicolumn{1}{c|}{$f_{mqq}^2/4\pi$} & $\Lambda_m$ 
&\multicolumn{2}{c|}{$\sigma=0.06~\text{GeV}^2$}
&\multicolumn{2}{c|}{$\sigma=0.09~\text{GeV}^2$}
&\multicolumn{2}{c|}{$\sigma=0.12~\text{GeV}^2$}\\
\cline{6-11}
&&&&
&\multicolumn{1}{c|}{ $m_{eff}$} & \multicolumn{1}{c|}{$g_{mqq}^2/4\pi$} 
&\multicolumn{1}{c|}{ $m_{eff}$} & \multicolumn{1}{c|}{$g_{mqq}^2/4\pi$} 
& \multicolumn{1}{c|}{$m_{eff}$} & \multicolumn{1}{c|}{$g_{mqq}^2/4\pi$}\\
\hline\hline
$\pi$    & 139 &  93 & 0.0444 &  678 & 206 & 0.39 & 254 & 0.59 & 294 & 0.79\\
$\kappa$ & 494 & 113 & 0.380  &  966 & 266 & 0.44 & 312 & 0.61 & 351 & 0.76\\
$\eta$   & 547 & 112 & 0.475  & 1008 & 266 & 0.45 & 312 & 0.62 & 351 & 0.78\\
\hline
\end{tabular}
\end{table}

\newpage

%%++ Table:3
\begin{table}% Sigma = 0.06 GeV^2
\setcaptionmargin{0mm}
\onelinecaptionsfalse
\captionstyle{flushleft}
\caption{ The mass is calculated for 
the baryon multiplet for the string tension 
$\sigma=0.06~(\text{GeV}^2)$. The $C_0$ is
adjusted such as to yield the correct value for the nucleon. 
The contributions of the exchange potentials are also shown.
All numbers are in units of MeV.
\label{tab:spectrum0}}
\bigskip
\begin{tabular}{|c|l|r|r|r|r|r|r|r|c|}
\hline
& N & $V_{\rm coul}$ & $V_{\rm hf}$ & $V_{\pi}$ & $V_\kappa$ & $V_\eta$ & $M_N$ & $M_{\text exp}$ & $C_0$ \\
\hline\hline
PS   & p,n        & -130 &  -9 & -175 &   0 &   4 &  939 &  939 & 174 \\
     & $\Lambda$  & -138 &  -9 & -105 & -25 &   4 & 1121 & 1116 & \\
     & $\Sigma$   & -138 &  -7 &  -12 & -42 & -12 & 1184 & 1193 & \\
     & $\Xi$      & -147 &  -8 &    0 & -42 & -15 & 1327 & 1318 & \\
\cline{2-9}
     & $\Delta$   & -130 &   9 &  -35 &   0 &  -4 & 1089 & 1232 & \\
     & $\Sigma^*$ & -138 &   8 &  -12 & -17 &   4 & 1240 & 1385 & \\
     & $\Xi^*$    & -147 &   7 &    0 & -17 &   0 & 1383 & 1516 & \\
     & $\Omega$   & -158 &   7 &    0 &   0 & -14 & 1518 & 1672 & \\
\hline\hline
PV   & p,n        & -130 &  -9 & -121 &   0 &   1 &  939 &  939 & 157 \\
     & $\Lambda$  & -138 &  -9 &  -72 & -19 &   1 & 1106 & 1116 & \\
     & $\Sigma$   & -138 &  -7 &   -8 & -32 &  -7 & 1151 & 1193 & \\
     & $\Xi$      & -147 &  -8 &    0 & -32 & -13 & 1288 & 1318 & \\
\cline{2-9}
     & $\Delta$   & -130 &   9 &  -24 &   0 &  -1 & 1052 & 1232 & \\
     & $\Sigma^*$ & -138 &   8 &   -8 & -13 &   3 & 1195 & 1385 & \\
     & $\Xi^*$    & -147 &   7 &    0 & -13 &  -3 & 1332 & 1516 & \\
     & $\Omega$   & -158 &   7 &    0 &   0 & -20 & 1462 & 1672 & \\
\hline
\end{tabular}
\end{table}

\newpage

%%++ Table:4
\begin{table}% Sigma=0.09 GeV^2
\setcaptionmargin{0mm}
\onelinecaptionsfalse
\captionstyle{flushleft}
\caption{ The mass is calculated for 
the baryon multiplet for the string tension 
$\sigma=0.09~(\text{GeV}^2)$. The $C_0$ is
adjusted such as to yield the correct value for the nucleon. 
The contributions of the exchange potentials are also shown.
All numbers are in units of MeV.
\label{tab:spectrum1}}
\bigskip
\begin{tabular}{|c|l|r|r|r|r|r|r|r|c|}
\hline
& N & $V_{\rm coul}$ & $V_{\rm hf}$ & $V_{\pi}$ & $V_\kappa$ & $V_\eta$ & $M_N$ & $M_{\text exp}$ & $C_0$ \\
\hline\hline
PS   & p,n        & -159 & -11 & -314 &   0 &   9 &  939 &  939 & 175 \\
     & $\Lambda$  & -167 & -11 & -189 & -52 &   9 & 1148 & 1116 & \\
     & $\Sigma$   & -167 &  -9 &  -21 & -86 & -24 & 1250 & 1193 & \\
     & $\Xi$      & -177 & -10 &    0 & -86 & -31 & 1396 & 1318 & \\
\cline{2-9}
     & $\Delta$   & -159 &  11 &  -63 &   0 &  -9 & 1195 & 1232 & \\
     & $\Sigma^*$ & -167 &  10 &  -21 & -34 &   8 & 1353 & 1385 & \\
     & $\Xi^*$    & -177 &   9 &    0 & -34 &   1 & 1499 & 1516 & \\
     & $\Omega$   & -187 &   8 &    0 &   0 & -30 & 1634 & 1672 & \\
\hline\hline
PV   & p,n        & -159 & -11 & -222 &   0 &   3 &  939 &  939 & 146 \\
     & $\Lambda$  & -167 & -11 & -133 & -39 &   3 & 1123 & 1116 & \\
     & $\Sigma$   & -167 &  -9 &  -15 & -65 & -16 & 1199 & 1193 & \\
     & $\Xi$      & -177 & -10 &    0 & -65 & -27 & 1335 & 1318 & \\
\cline{2-9}
     & $\Delta$   & -159 &  11 &  -44 &   0 &  -3 & 1132 & 1232 & \\
     & $\Sigma^*$ & -167 &  10 &  -15 & -26 &   6 & 1279 & 1385 & \\
     & $\Xi^*$    & -177 &   9 &    0 & -26 &  -5 & 1415 & 1516 & \\
     & $\Omega$   & -187 &   8 &    0 &   0 & -37 & 1539 & 1672 & \\
\hline
\end{tabular}
\end{table}

\newpage

%%++ Table:5
\begin{table}% Sigma=0.12 GeV^2
\setcaptionmargin{0mm}
\onelinecaptionsfalse
\captionstyle{flushleft}
\caption{ The mass is calculated for 
the baryon multiplet for the string tension 
$\sigma=0.12~(\text{GeV}^2)$. The $C_0$ is
adjusted such as to yield the correct value for the nucleon. 
The contributions of the exchange potentials are also shown.
All numbers are in units of MeV.
\label{tab:spectrum1b}}
\bigskip
\begin{tabular}{|c|l|r|r|r|r|r|r|r|c|}
\hline
& N & $V_{\rm coul}$ & $V_{\rm hf}$ & $V_{\pi}$ & $V_\kappa$ & $V_\eta$ & $M_N$ & $M_{\text exp}$ & $C_0$ \\
\hline\hline
PS   & p,n        & -182 & -13 & -458 &   0 &  14 &  939 &  939 & 184 \\
     & $\Lambda$  & -191 & -13 & -275 & -82 &  14 & 1174 & 1116 & \\
     & $\Sigma$   & -191 & -11 &  -31 &-136 & -39 & 1312 & 1193 & \\
     & $\Xi$      & -200 & -12 &    0 &-136 & -51 & 1462 & 1318 & \\
\cline{2-9}
     & $\Delta$   & -182 &  13 &  -92 &   0 & -14 & 1303 & 1232 & \\
     & $\Sigma^*$ & -191 &  12 &  -31 & -55 &  13 & 1468 & 1385 & \\
     & $\Xi^*$    & -200 &  11 &    0 & -55 &   1 & 1618 & 1516 & \\
     & $\Omega$   & -210 &  10 &    0 &   0 & -48 & 1753 & 1672 & \\
\hline\hline
PV   & p,n        & -182 & -13 & -327 &   0 &   6 &  939 &  939 & 143 \\
     & $\Lambda$  & -191 & -13 & -196 & -61 &   6 & 1142 & 1116 & \\
     & $\Sigma$   & -191 & -11 &  -22 &-101 & -25 & 1247 & 1193 & \\
     & $\Xi$      & -200 & -12 &    0 &-101 & -42 & 1383 & 1318 & \\
\cline{2-9}
     & $\Delta$   & -182 &  13 &  -65 &   0 &  -6 & 1215 & 1232 & \\
     & $\Sigma^*$ & -191 &  12 &  -22 & -40 &  10 & 1365 & 1385 & \\
     & $\Xi^*$    & -200 &  11 &    0 & -40 &  -7 & 1501 & 1516 & \\
     & $\Omega$   & -210 &  10 &    0 &   0 & -56 & 1622 & 1672 & \\
\hline
\end{tabular}
\end{table}

\newpage

%%++ Table:6
\begin{table}% Sigma=0.06 GeV^2
\setcaptionmargin{0mm}
\onelinecaptionsfalse
\captionstyle{flushleft}
\caption{ The result on the mass in units of MeV and the magnetic moment
in units of the nuclear magneton of the multichannel calculation where
the single-particle orbitals (spo's) $0\bigl(\tfrac12\bigr)^{++}$, 
$0\bigl(\tfrac12\bigr)^{-+}$, $0\bigl(\tfrac32\bigr)^{++}$ and either 
$1\bigl(\tfrac12\bigr)^{++}$ (PS) or $0\bigl(\tfrac52\bigr)^{++}$ (PV)
are used as input. 
The string tension is 
fixed at $\sigma=0.06~(\text{GeV})^2$, the $C_0$ is adjusted to yield
the correct nucleon mass, in case of the PS-coupling $C_0(1)=174~\text{MeV}$,
$C_0(4)=204~\text{MeV}$ and in case of the PV-coupling $C_0(1)=157~\text{MeV}$,
$C_0(4)=182~\text{MeV}$. 
\label{result4channel0}}
\bigskip
\begin{tabular}{|l||r|d|r|d||r|d|r|d||r|d|}
\hline
N & \multicolumn{4}{c||}{PS-coupling} &  \multicolumn{4}{c||}{PV-coupling} & \multicolumn{2}{c|}{exp}\\
\cline{2-9}
& \multicolumn{2}{c|}{1 spo} &  \multicolumn{2}{c||}{4 spo's} & 
\multicolumn{2}{c|}{1 spo} &  \multicolumn{2}{c||}{4 spo's} &  \multicolumn{2}{c|}{} \\
\cline{2-11}
& $m_N$ & \multicolumn{1}{c|}{$\mu_N$} & $m_N$ & \multicolumn{1}{c||}{$\mu_N$} 
& $m_N$ & \multicolumn{1}{c|}{$\mu_N$} & $m_N$ & \multicolumn{1}{c||}{$\mu_N$} 
& $m_N$ & \multicolumn{1}{c|}{$\mu_N$} \\
\hline\hline
p 		&  939 &  3.58 &  939 &  3.18 &  939 &  3.64 &  939 &  3.19 &  938 &  2.79 \\
n 		&  939 & -2.44 &  939 & -2.15 &  939 & -2.50 &  939 & -2.17 &  940 & -1.91 \\  
$\Lambda$	& 1121 & -0.78 & 1133 & -0.75 & 1106 & -0.78 & 1120 & -0.77 & 1116 & -0.61 \\  
$\Sigma^{+}$	& 1184 &  3.43 & 1221 &  3.15 & 1151 &  3.48 & 1193 &  3.26 & 1189 &  2.46 \\  
$\Sigma^{0}$	& 1184 &  1.02 & 1221 &  0.94 & 1151 &  1.02 & 1193 &  0.96 & 1193 & \\  
$\Sigma^{-}$	& 1184 & -1.39 & 1221 & -1.28 & 1151 & -1.43 & 1193 & -1.34 & 1197 & -1.16 \\  
$\Xi^{0}$	& 1327 & -1.83 & 1363 & -1.68 & 1288 & -1.84 & 1331 & -1.78 & 1315 & -1.25 \\  
$\Xi^{-}$	& 1327 & -0.62 & 1363 & -0.63 & 1288 & -0.62 & 1331 & -0.62 & 1321 & -0.65 \\  
\hline\hline
$\Delta^{++}$	& 1089 &  7.13 & 1157 &  6.50 & 1052 &  7.24 & 1112 &  6.72 & 1232 &  4.52 \\  
$\Delta^{+}$	& 1089 &  3.52 & 1157 &  3.21 & 1052 &  3.56 & 1112 &  3.31 & 1232 & \\
$\Delta^{0}$	& 1089 & -0.09 & 1157 & -0.08 & 1052 & -0.13 & 1112 & -0.11 & 1232 & \\
$\Delta^{-}$	& 1089 & -3.70 & 1157 & -3.37 & 1052 & -3.81 & 1112 & -3.53 & 1232 & \\
$\Sigma^{+*}$	& 1240 &  3.98 & 1311 &  3.68 & 1195 &  4.05 & 1258 &  3.84 & 1383 & \\
$\Sigma^{0*}$	& 1240 &  0.37 & 1311 &  0.31 & 1195 &  0.36 & 1258 &  0.33 & 1384 & \\
$\Sigma^{-*}$	& 1240 & -3.70 & 1311 & -3.05 & 1195 & -3.32 & 1258 & -3.18 & 1387 & \\
$\Xi^{0*}$	& 1383 &  0.83 & 1456 &  0.77 & 1332 &  0.85 & 1395 &  0.82 & 1532 & \\
$\Xi^{-*}$	& 1383 & -2.79 & 1456 & -2.68 & 1332 & -2.83 & 1395 & -2.76 & 1535 & \\
$\Omega^{-}$	& 1518 & -2.33 & 1590 & -2.18 & 1462 & -2.34 & 1522 & -2.20 & 1672 & -2.02 \\  
\hline
\end{tabular}
\end{table}

\newpage

%%++ Table:7
\begin{table}% Sigma=0.09 GeV^2
\setcaptionmargin{0mm}
\onelinecaptionsfalse
\captionstyle{flushleft}
\caption{ The result on the mass in units of MeV and the magnetic moment
in units of the nuclear magneton of the multichannel calculation where
the single-particle orbitals (spo's) $0\bigl(\tfrac12\bigr)^{++}$, 
$0\bigl(\tfrac12\bigr)^{-+}$, $0\bigl(\tfrac32\bigr)^{++}$ and either 
$1\bigl(\tfrac12\bigr)^{++}$ (PS) or $0\bigl(\tfrac52\bigr)^{++}$ (PV)
are used as input. 
The string tension is 
fixed at $\sigma=0.09~(\text{GeV})^2$, the $C_0$ is adjusted to yield
the correct nucleon mass, in case of the PS-coupling $C_0(1)=175~\text{MeV}$,
$C_0(4)=221~\text{MeV}$ and in case of the PV-coupling $C_0(1)=146~\text{MeV}$,
$C_0(4)=190~\text{MeV}$. 
\label{result4channel}}
\bigskip
\begin{tabular}{|l||r|d|r|d||r|d|r|d||r|d|}
\hline
N & \multicolumn{4}{c||}{PS-coupling} &  \multicolumn{4}{c||}{PV-coupling} & \multicolumn{2}{c|}{exp}\\
\cline{2-9}
& \multicolumn{2}{c|}{1 spo} &  \multicolumn{2}{c||}{4 spo's} & 
\multicolumn{2}{c|}{1 spo} &  \multicolumn{2}{c||}{4 spo's} & \multicolumn{2}{c|}{} \\
\cline{2-11}
& $m_N$ & \multicolumn{1}{c|}{$\mu_N$} & $m_N$ & \multicolumn{1}{c||}{$\mu_N$} 
& $m_N$ & \multicolumn{1}{c|}{$\mu_N$} & $m_N$ & \multicolumn{1}{c||}{$\mu_N$} 
& $m_N$ & \multicolumn{1}{c|}{$\mu_N$} \\
\hline\hline
p 		&  939 &  2.99 &  939 &  2.65 &  939 &  3.07 &  939 &  2.61 &  938 &  2.79 \\
n 		&  939 & -2.06 &  939 & -1.80 &  939 & -2.13 &  939 & -1.80 &  940 & -1.91 \\  
$\Lambda$	& 1148 & -0.69 & 1160 & -0.67 & 1123 & -0.70 & 1146 & -0.69 & 1116 & -0.61 \\  
$\Sigma^{+}$	& 1250 &  2.88 & 1292 &  2.60 & 1199 &  2.94 & 1264 &  2.65 & 1189 &  2.46 \\  
$\Sigma^{0}$	& 1250 &  0.85 & 1292 &  0.77 & 1199 &  0.86 & 1264 &  0.78 & 1193 & \\  
$\Sigma^{-}$	& 1250 & -1.17 & 1292 & -1.06 & 1199 & -1.22 & 1264 & -1.10 & 1197 & -1.16 \\  
$\Xi^{0}$	& 1396 & -1.58 & 1435 & -1.43 & 1335 & -1.61 & 1402 & -1.51 & 1315 & -1.25 \\  
$\Xi^{-}$	& 1396 & -0.57 & 1435 & -0.57 & 1335 & -0.57 & 1402 & -0.57 & 1321 & -0.65 \\  
\hline\hline
$\Delta^{++}$	& 1195 &  5.95 & 1297 &  5.35 & 1132 &  6.08 & 1240 &  5.43 & 1232 &  4.52 \\  
$\Delta^{+}$	& 1195 &  2.92 & 1297 &  2.63 & 1132 &  2.96 & 1240 &  2.70 & 1232 & \\
$\Delta^{0}$	& 1195 & -0.11 & 1297 & -0.10 & 1132 & -0.15 & 1240 & -0.13 & 1232 & \\
$\Delta^{-}$	& 1195 & -3.15 & 1297 & -2.82 & 1132 & -3.27 & 1240 & -2.96 & 1232 & \\
$\Sigma^{+*}$	& 1353 &  3.28 & 1460 &  2.98 & 1279 &  3.36 & 1391 &  3.13 & 1383 & \\
$\Sigma^{0*}$	& 1353 &  0.24 & 1460 &  0.19 & 1279 &  0.24 & 1391 &  0.21 & 1384 & \\
$\Sigma^{-*}$	& 1353 & -2.79 & 1460 & -2.59 & 1279 & -2.88 & 1391 & -2.71 & 1387 & \\
$\Xi^{0*}$	& 1499 &  0.60 & 1609 &  0.54 & 1415 &  0.63 & 1527 &  0.60 & 1532 & \\
$\Xi^{-*}$	& 1499 & -2.43 & 1609 & -2.32 & 1415 & -2.49 & 1527 & -2.41 & 1535 & \\
$\Omega^{-}$	& 1634 & -2.07 & 1743 & -1.92 & 1539 & -2.09 & 1649 & -1.93 & 1672 & -2.02 \\  
\hline
\end{tabular}
\end{table}

\newpage

%%++ Table:8
\begin{table}% Sigma=0.12 GeV^2
\setcaptionmargin{0mm}
\onelinecaptionsfalse
\captionstyle{flushleft}
\caption{ The result on the mass in units of MeV and the magnetic moment
in units of the nuclear magneton of the multichannel calculation where
the single-particle orbitals (spo's) $0\bigl(\tfrac12\bigr)^{++}$, 
$0\bigl(\tfrac12\bigr)^{-+}$, $0\bigl(\tfrac32\bigr)^{++}$ and 
either $1\bigl(\tfrac12\bigr)^{++}$ (PS) or $0\bigl(\tfrac52\bigr)^{++}$ (PV)
are used as input. 
The string tension is 
fixed at $\sigma=0.12~(\text{GeV})^2$, the $C_0$ is adjusted to yield
the correct nucleon mass, in case of the PS-coupling $C_0(1)=184~\text{MeV}$,
$C_0(4)=242~\text{MeV}$ and in case of the PV-coupling $C_0(1)=143~\text{MeV}$,
$C_0(4)=204~\text{MeV}$.
\label{result4channelb}}
\bigskip
\begin{tabular}{|l||r|d|r|d||r|d|r|d||r|d|}
\hline
N & \multicolumn{4}{c||}{PS-coupling} &  \multicolumn{4}{c||}{PV-coupling} & \multicolumn{2}{c|}{exp}\\
\cline{2-9}
& \multicolumn{2}{c|}{1 spo} &  \multicolumn{2}{c||}{4 spo's} & 
\multicolumn{2}{c|}{1 spo} &  \multicolumn{2}{c||}{4 spo's} & \multicolumn{2}{c|}{} \\
\cline{2-11}
& $m_N$ & \multicolumn{1}{c|}{$\mu_N$} & $m_N$ & \multicolumn{1}{c||}{$\mu_N$} 
& $m_N$ & \multicolumn{1}{c|}{$\mu_N$} & $m_N$ & \multicolumn{1}{c||}{$\mu_N$} 
& $m_N$ & \multicolumn{1}{c|}{$\mu_N$} \\
\hline\hline
p 		&  939 &  2.66 &  939 &  2.35 &  939 &  2.74 &  939 &  2.27 &  938 &  2.79 \\  
n 		&  939 & -1.85 &  939 & -1.62 &  939 & -1.93 &  939 & -1.58 &  940 & -1.91 \\  
$\Lambda$	& 1174 & -0.64 & 1184 & -0.61 & 1142 & -0.65 & 1169 & -0.63 & 1116 & -0.61 \\  
$\Sigma^{+}$	& 1312 &  2.56 & 1350 &  2.29 & 1247 &  2.63 & 1325 &  2.32 & 1189 &  2.46 \\  
$\Sigma^{0}$	& 1312 &  0.75 & 1350 &  0.67 & 1247 &  0.76 & 1325 &  0.67 & 1193 & \\
$\Sigma^{-}$	& 1312 & -1.05 & 1350 & -0.94 & 1247 & -1.11 & 1325 & -0.97 & 1197 & -1.16 \\  
$\Xi^{0}$	& 1462 & -1.44 & 1495 & -1.29 & 1382 & -1.46 & 1463 & -1.37 & 1315 & -1.25 \\  
$\Xi^{-}$	& 1462 & -0.53 & 1495 & -0.53 & 1382 & -0.53 & 1463 & -0.53 & 1321 & -0.65 \\  
\hline\hline
$\Delta^{++}$	& 1303 &  5.28 & 1430 &  4.72 & 1214 &  5.43 & 1364 &  4.85 & 1232 &  4.52 \\  
$\Delta^{+}$	& 1303 &  2.57 & 1430 &  2.30 & 1214 &  2.63 & 1364 &  2.35 & 1232 & \\
$\Delta^{0}$	& 1303 & -0.13 & 1430 & -0.11 & 1214 & -0.18 & 1364 & -0.15 & 1232 & \\
$\Delta^{-}$	& 1303 & -2.84 & 1430 & -2.53 & 1214 & -2.98 & 1364 & -2.65 & 1232 & \\
$\Sigma^{+*}$	& 1468 &  2.88 & 1601 &  2.60 & 1364 &  2.97 & 1520 &  2.73 & 1383 & \\
$\Sigma^{0*}$	& 1468 &  0.18 & 1601 &  0.13 & 1364 &  0.17 & 1520 &  0.14 & 1384 & \\
$\Sigma^{-*}$	& 1468 & -2.53 & 1601 & -2.34 & 1364 & -2.63 & 1520 & -2.45 & 1387 & \\
$\Xi^{0*}$	& 1618 &  0.49 & 1754 &  0.43 & 1500 &  0.52 & 1657 &  0.48 & 1532 & \\
$\Xi^{-*}$	& 1618 & -2.22 & 1754 & -2.10 & 1500 & -2.28 & 1657 & -2.20 & 1535 & \\
$\Omega^{-}$	& 1753 & -1.91 & 1887 & -1.75 & 1621 & -1.94 & 1774 & -1.76 & 1672 & -2.02 \\  
\hline
\end{tabular}
\end{table}

\newpage

%%++ Table:9
\begin{table}
\setcaptionmargin{0mm}
\onelinecaptionsfalse
\captionstyle{flushleft}
\caption{ The anomalous magnetic moment of the quark in units
of the nucleon magneton for different string tensions $\sigma$. 
The parameters are taken from Table~\ref{tab:parameters}.
The first set is the prediction for only the pion loops, while 
the second set is with both pion and kaon loops included and the 
third set shows the results where pion,
kaon and eta loops are taken into account.
\label{tab:anommagnmom}}
\bigskip
\begin{tabular}{|l|d|d|d|d|d|d|}
\hline
$\kappa_{f_i}$ & 
\multicolumn{2}{c|}{$\sigma=0.06~\text{GeV}^2$} &
\multicolumn{2}{c|}{$\sigma=0.09~\text{GeV}^2$} &
\multicolumn{2}{c|}{$\sigma=0.12~\text{GeV}^2$}\\
\cline{2-7}
& \multicolumn{1}{c|}{PS}& \multicolumn{1}{c|}{PV}
& \multicolumn{1}{c|}{PS}& \multicolumn{1}{c|}{PV}
& \multicolumn{1}{c|}{PS}& \multicolumn{1}{c|}{PV}\\
\hline\hline
& \multicolumn{6}{c|}{Pion loops}\\
\cline{2-7}
$\kappa_u$ &  0.065 &  0.091 &  0.089 &  0.121 &  0.109 &  0.147\\
$\kappa_d$ & -0.110 & -0.153 & -0.141 & -0.193 & -0.166 & -0.224\\
$\kappa_s$ &  0.0   &  0.0   &  0.0   &  0.0   &  0.0   &  0.0 \\
\hline\hline
& \multicolumn{6}{c|}{Pion and kaon loops}\\
\cline{2-7}
$\kappa_u$ &  0.097 &  0.136 &  0.129 &  0.178 &  0.155 &  0.211\\
$\kappa_d$ & -0.100 & -0.138 & -0.129 & -0.176 & -0.152 & -0.206\\
$\kappa_s$ & -0.028 & -0.040 & -0.039 & -0.054 & -0.048 & -0.066\\
\hline\hline
& \multicolumn{6}{c|}{Pion, kaon and eta loops}\\
\cline{2-7}
$\kappa_u$ &  0.092 &  0.131 &  0.123 &  0.172 &  0.148 &  0.205\\
$\kappa_d$ & -0.097 & -0.136 & -0.126 & -0.173 & -0.149 & -0.202\\
$\kappa_s$ & -0.021 & -0.025 & -0.031 & -0.038 & -0.038 & -0.048\\
\hline
\end{tabular}
\end{table}
\end{document}